%% file: HEXD_Stress_Recovery.tex
\documentclass{article}   

\usepackage{graphicx}
\usepackage{hyperref}
\hypersetup{
    colorlinks=true,
    linkcolor=blue,
    urlcolor=blue,
    citecolor=blue,
}

\usepackage{subcaption}

\usepackage{float}
\usepackage{amssymb}
\usepackage{xfrac}
\usepackage{miller}

\usepackage{siunitx}
\usepackage{mathtools}
\usepackage{booktabs} 	
\usepackage[title]{appendix}
\input{mymathmacros}

\begin{document}

\title{Using Discrete Harmonic Expansions and Equilibrium Conditions\\ to Estimate Intragrain Stress Distributions in Polycrystals\\ from Grain-Averaged Data}

\author{Paul R. Dawson \and Matthew P. Miller}

\date{Sibley School of Mechanical and Aerospace Engineering\\ Cornell University\\  Ithaca, NY 14850, USA \\ \today}

\maketitle

\begin{abstract}
The application of harmonic expansions to estimate intra-grain stress distributions from grain-averaged stress data is presented that extends the capabilities of the open source code, \mechmonics.
The method is based on using an optimization algorithm to determine the harmonic expansion weights that reduce the violation of equilibrium  while maintaining prescribed grain-averages.  The method is demonstrated using synthetic data generated for uniaxial extension of a virtual polycrystal with the \mechmet\, code.

\end{abstract}

\section{Introduction}

Discrete harmonic expansions provide a mechanism to determine a reduced-order representation of stress distributions within grains.  
A reduction of  two or more orders of magnitude was demonstrated in \cite{Mechmonics_2021}  while retaining essential features of spatial variations in stress distributions generated by means of finite element simulations.  
An application is presented here for using the expansions to estimate intra-grain stress distributions  from data that provides only grain-averaged values of stress.  
Such data can be generated by High Energy x-ray Diffraction Microscopy (HEDM) experiments.  
The intra-grain details of the distributions are not captured with the current far-field HEDM methods, but are 
of considerable interest to researchers.  
Together with the harmonic expansion representations the stress,  the imposition of equilibrium conditions provides supplemental requirements that enable estimation of the intra-grain stress variations.

The methodology consists of (1) representing stresses using an expansion of known modes but unknown weights; (2) writing equilibrium conditions at the grain boundaries and grain interiors in terms of the expansions; and (3) using an optimization methodology to determine the weights that allow the stress distributions to best satisfy those conditions.  
The method is implemented in \matlab\cite{matlab}, and requires as input the 
harmonic modes generated by \mechmonics.

The paper first lays out the mechanics foundations of the approach and its numerical implementation.  
It then presents a demonstration example of tensile loading of a virtual polycrystal.    
 The stress distribution data consist of grain-averaged stresses derived from a finite element simulation, which mimic data that are extracted from the HEDM experiments.
 Using the method, a stress distribution having intra-granular variations  is reconstructed and
 compared to the original distribution.
 Explanation is offered for why the reconstructed distribution, while decreasing equilibrium violations, does not reproduce the original distribution.

\section{Recovery of Intra-Grain Stress Distributions}
\label{sec:mode-based_stress_recovery}

\subsection{Overview}
Experimental data has limited spatial resolution either from physical attributes of the equipment or practical constraints in the data collection.  
An ever-present challenge is to fill in the picture that the data portrays incompletely or 
in a self-contradicting manner. 
 Experimental data in the form of grain-averaged quantities, like elastic strain or stress, presents a contradiction at grain boundaries in the form of imbalanced surface tractions at grain boundaries.  
 In polycrystals, stress distributions within an individual grain must vary spatially to facilitate traction equilibrium at the boundaries with other grains.  
Far Field HEDM experiments are capable of estimating the average stress over each grain within an aggregate, with the consequence that the spatial distribution of stress within grains is lost.  
It is possible to estimate the spatial variations, however, given the topology of the polycrystal, which can be obtained from near field HEDM data, together with the average quantities by modifying the distribution on the basis of rectifying the violations in equilibrium.  

The methodology  proposed here is presented in the following order:
\begin{enumerate}
\item In this section the pertinent equilibrium relations are summarized.
\item In Section~\ref{sec:harmonics_background}, background is provided on harmonic expansions, which form the backbone of the method by providing a reduced-order representation of the intra-grain stress distributions.
 \item In Section~\ref{sec:fe_methodology}, the matrix formulation leading to the equilibrium-based objective function is presented.  The formulation facilitates the calling of an in-line optimization algorithm.  
 \item In Section~\ref{sec:demo_application}, the methodology is demonstrated for a stress distribution generated with \mechmet~\cite{mechmet_immi}.   Critiquing the reconstructed stress distribution is then possible through comparison between the reconstructed distribution and one available from harmonic expansion representation of the mechmet solution. 
 \item In Section~\ref{sec:discussion} some implications of basing the objective function on equilibrium, as opposed to compatibility, are discussed.
\end{enumerate}

\subsection{Objective function based on equilibrium conditions}

The purpose of the methodology presented here is to compute plausible intra-grain spatial variations of stress distributions for which we only have grain averages.  
We turn to equilibrium requirements because the averages are associated with 
polycrystals that are loaded (quasi-)statically and, consequently,
variations that improve upon equilibrium conditions not only are plausible, but
are desirable.  
Foundational theory of solid mechanics provides multiple expressions of equilibrium~\cite{bower_book}.
Commonly employed are a local form written in terms of the divergence of the stress at points within the domain's volume and a global form written in terms of the resultant of tractions integrated over a domain's surface.   A variant of the latter is to require that the traction vectors at points along an arbitrary parting surface must be equal in magnitude and opposite in sign across the opposing sides.  
This expression of equilibrium is useful for polycrystals and is applied here for grain boundaries.  In addition, the local form of equilibrium is applied.  

\leftline{\it Grain boundary interfaces:}

\begin{figure}[htbp]
	\centering		
	\begin{subfigure}{.2\textwidth}
		\centering
		\includegraphics[width=1\linewidth]{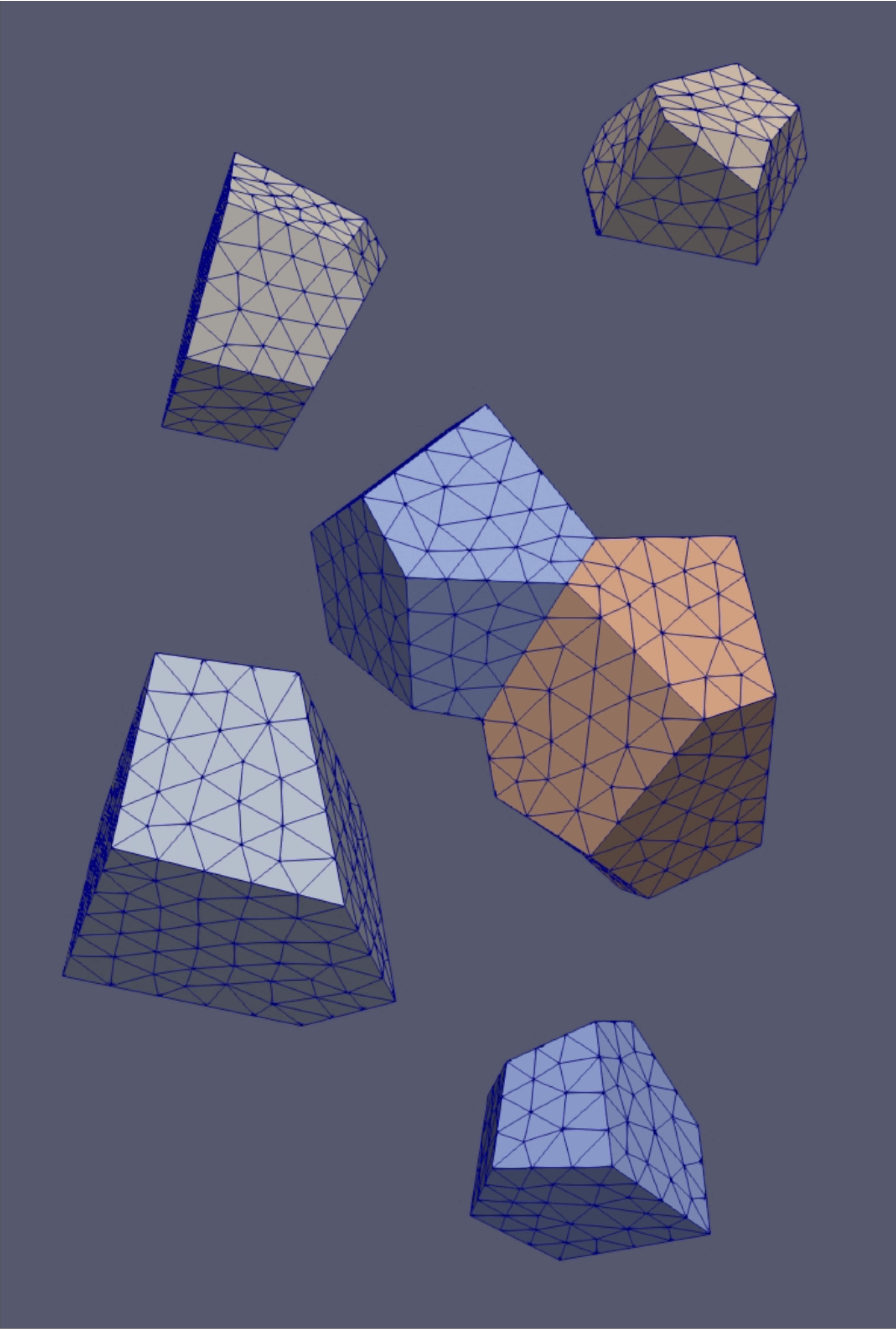}
		\caption{ }
		\label{fig:traction_constraint_1}
	\end{subfigure}%
	\quad
		\begin{subfigure}{.4\textwidth}
		\centering
		\includegraphics[width=1\linewidth]{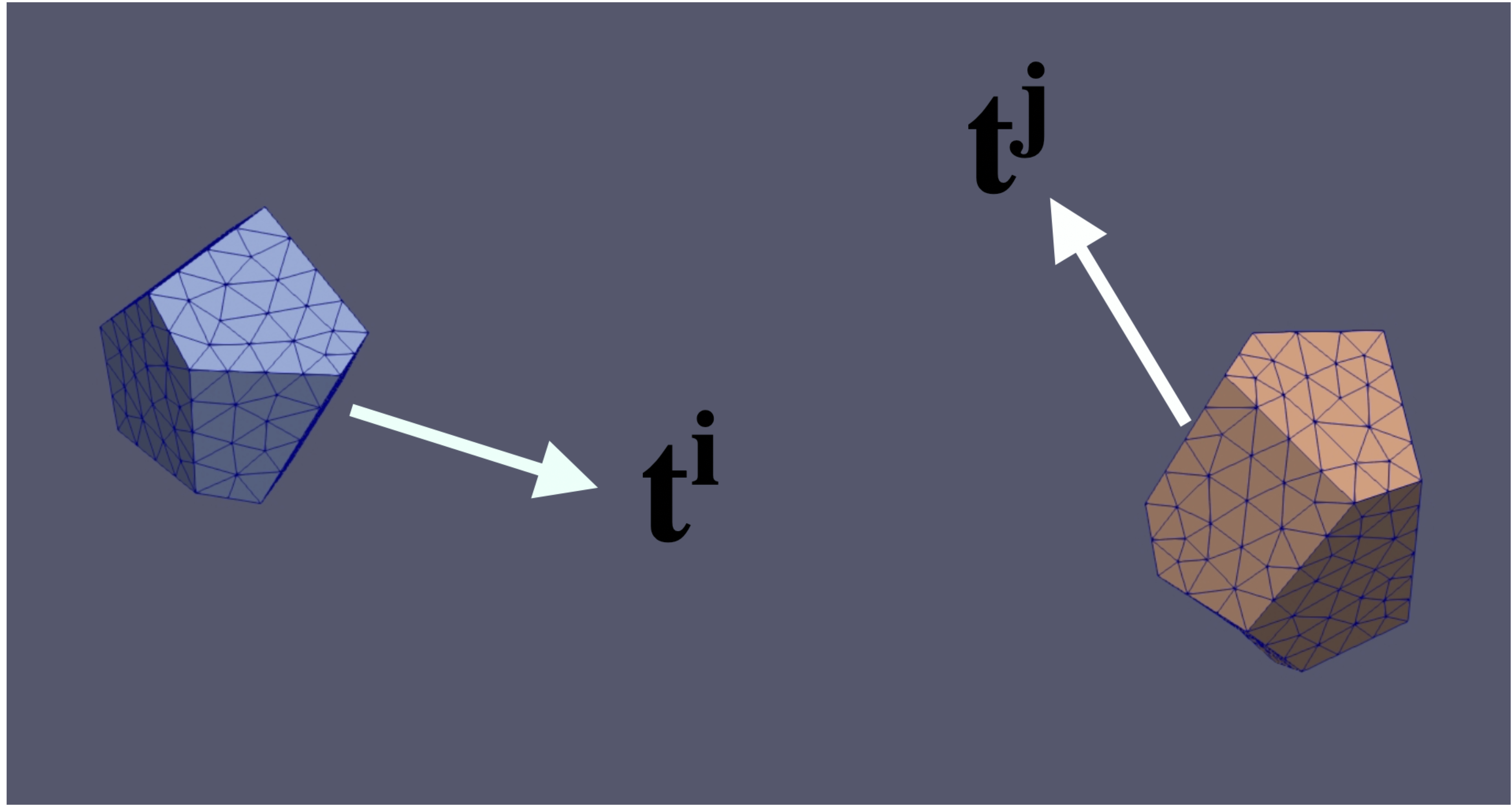}
		\caption{ }
		\label{fig:traction_constraint_2}
	\end{subfigure}
		\caption{a) Neighboring grain pair (b) opposing traction vectors at one point of the interface.}
		\label{fig:traction_constraint}
\end{figure}

At grain boundaries the forces transmitted across  interfaces must be balanced.  
The Cauchy formula:
\begin{equation}
 \vctr{t} = \cauchy  \cdot \vctr{n}              
  \label{eq:cauchy_formula}
\end{equation}
provides the pertinent relation between the traction on a surface, $\vctr{t}$, the surface normal, $\vctr{n}$, 
and the Cauchy stress, $\cauchy$.
the force balance at any point along adjoining surfaces of two grains (see Figure~\ref{fig:traction_constraint}) can be written with the traction vectors of the
adjoining grains as:
\begin{equation}
\vctr{t}^i + \vctr{t}^j  = 0
                \label{eq:interface_equil}
\end{equation}
If Equation~\ref{eq:interface_equil} is not satisfied, an equilibrium violation, $\vctr{f}^b$, exists:
\begin{equation}
\vctr{f}^b = \vctr{t}^i + \vctr{t}^j  
                \label{eq:boundary_resultant}
\end{equation}
 $\vctr{f}^b$ is a (vector) force imbalance per unit area at the adjoining point.

\leftline{\it Grain interiors:}

\begin{figure}[htbp]
		\centering
		\includegraphics[width=0.5\linewidth]{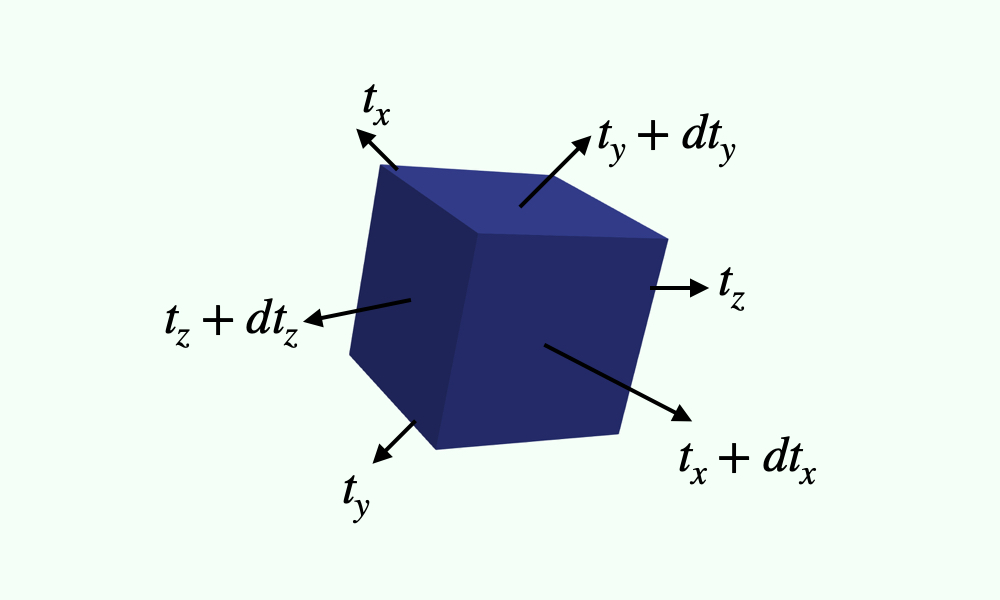}
		\caption{ Intra-grain Infinitessimal cube showing active tractions on surfaces in the presence of a stress gradient.}
		\label{fig:equilibrium_constraint}
\end{figure}

Local form of equilibrium, expressed in term of gradients of the stress tensor, is written in the familiar form:
\begin{equation}
\nabla \tnsr{ \cauchy} = 0
                \label{eq:div_sigma}
\end{equation}
assuming the body forces are negligible.  
This equation is derived by considering force balances  over an infinitesimal cube in the presence of a spatially varying stress distribution, shown in Figure~\ref{fig:equilibrium_constraint} in terms of tractions.
If Equation~\ref{eq:div_sigma} is not satisfied, an equilibrium violation, $\vctr{f}^v$, exists:
\begin{equation}
\vctr{f}^v = \nabla \tnsr{ \cauchy}
                \label{eq:local_resultant}
\end{equation}
 $\vctr{f}^c$ is a (vector) force imbalance over a differential volume.

\leftline{\it Objective function:}

The equilbrium violation vectors are used to write the objective function.
From violations on the grain boundaries:
\begin{equation}
F^b = \vctr{f}^b \cdot  \vctr{f}^b
                \label{eq:boundary_residual}
\end{equation}
From violations within the grain interiors:
\begin{equation}
F^v = \vctr{f}^v \cdot  \vctr{f}^v
                \label{eq:volume_residual}
\end{equation}
$F^b$ and $F^v$ are combined to give an objective function that weights both measures of 
equilibrium violation:
\begin{equation}
F = \sum_{n^b}w^b F^b + \sum_{n^v}w^v F^v 
                \label{eq:total_residual}
\end{equation}
where $w^b$ and $w^v$ are weights given to the boundary and volume equilibrium violations, respectively, and $n^b$ and $n^v$ are the numbers of points at which those violations are determined.

\leftline{\it Data Constraint:}
The reconstructed stress distributions should conform to what is known about the distributions
from the experimental data.  Namely, that the average stress over a grain, given by:
\begin{equation}
\cauchy^{\rm{ave} }  =\frac{1}{V_{g}} \int_{V_g} \cauchy \dee{V}
                \label{eq:volume_average_stress}
\end{equation}
where $V_g$ is the grain volume so that:
\begin{equation}
\cauchy^{\rm{ave} } = \cauchy^{\rm{data}} 
                \label{eq:equality_constraint}
\end{equation}

\section{Background on Discrete Harmonic Expansions}
\label{sec:harmonics_background}

\leftline{\it Matrix notation for stress tensor}
For expedience in constructing matrix expressions needed in the computations, Voigt notation is employed here. 
Using this notation, the six independent components of the symmetric stress tensor are written as a vector: 
\begin{equation}
\{\Sigma\} = \left\{  
\begin{array}{c}
               \Sigma_1 \\
               \Sigma_2 \\
               \Sigma_3 \\
               \Sigma_4 \\
               \Sigma_5 \\ 
               \Sigma_6
 \end{array}
                \right\}
                = \left\{  
\begin{array}{c}
               \sigma_{xx} \\
               \sigma_{yy} \\
               \sigma_{zz} \\
               \sigma_{yz} \\
               \sigma_{xz} \\ 
               \sigma_{xy}
 \end{array}
                \right\}
                \label{eq:S_sixvector}
\end{equation}
The spatial distribution of each component  over a grain is expressed using 
expansion functions, $u^\alpha (\vctr{x})$:
\begin{equation}
 \Sigma_i({\vctr{x}})= \sum_k \beta^k_i u^k({\vctr{x}}) \hspace{1cm} (i = 1,n^c \,\,{\rm and}\,\, k = 1,n^m)
\label{eq:S_expansion}
\end{equation}
where $n^c$ is the number of stress components ($n^c \equiv6$), $n^m$ is the number of terms in the expansions and $\vctr{x}$ is the vector of coordinates of a point within the grain.
There is an expansion for every stress component within a grain.  
For any given grain, all of the stress components use the same expansion functions but every stress component has expansion coefficients that are independent of those associated with other stress components.
Correspondingly, the expansion coefficients, $\beta^k_i$, for one grain is $n^c \times n^m$.\\

\leftline{\it Evaluation of expansion functions:}
Discrete harmonic functions are chosen for the expansion functions.  
The individual expansion functions are referred to as modes and the associated
coefficients as weights.  
The grains of a polycrystal in general have different sizes and shapes.  
Because the modes depend on the domain size and shape, they differ from grain-to-grain.
The modes, therefore, must be computed specifically for each grain. 
This is accomplished by solving Laplace's equation
subject to the boundary condition of zero fluxes\cite{harmonics,wiki:laplace}.  
In the absence of essential boundary conditions, this system of equations is singular and thus has no unique solution.
Thus, we seek the singular values for the system given by:
\begin{equation}
	\nabla^2 u  -  \lambda u  = 0
	\label{eq:singularform}
\end{equation}
By performing a singular value decomposition (SVD),  
a set of singular values and vectors, $(\lambda^i , \{ U^i \} )$, may be extracted~\cite{wiki:svd}.
The singular vectors are the harmonic modes, or bases, of the expansion and are orthonormal.

The approach developed in \cite{Mechmonics_2021} is based on a finite element discretization of the polycrystal and representation of the modes using the finite element interpolation functions.
In this discretized system, there are a finite, rather than infinite, number of modes. 
The number of modes equals the number of nodes in the mesh for the grain.
With the finite element discretization, the modes are represented using the 
interpolation functions, $ [N({\bf{x}})]^e $ and nodal point values, $ \{ U\}^{k e} $, as:
\begin{equation}
u^k({\bf{x}}) = [ N({\bf{x}})]^e \{ U\}^{k e}  \hspace{1cm} (k = 1,n^m)
\label{eq:fem_rep_elemental}
\end{equation}
The stress components thus become:
\begin{equation}
 \Sigma_i({\bf{x}}) =  \sum_k \beta^k_i [ N({\bf{x}})] \{ U\} ^k
\label{eq:fem_rep_global}
\end{equation}
Here,  $ \{ U \}^k$ constitutes one of $k$ modes that quantify systematic spatial variations ranging from a constant value over the volume to more complex linear or quadratic distributions.  
The $\beta^k _i$ again are the weights.
The larger the relative value of one weight in comparison to others, the more significant
the functional dependence embodied in the corresponding mode.

The modes are determined {\it a priori}, so that when the expansion is substituted into 
Equation~\ref{eq:fem_rep_global}, the number of degrees of freedom in representing any one component of stress decreases from the 
number of finite element nodes to the number of spherical harmonic modes. 
This easily can be a factor greater than 10, and readily 100 or more.\\

\leftline{\it Evaluation of mode weights from known stress distributions}

Should a distribution over a grain already be known in terms of the nodal point values 
of the field variable, the expansion coefficients may be evaluated in the customary way for expansions with orthonormal modes.
To obtain $\beta^k$, the representation of $\{ \Sigma \} $ in Equation~\ref{eq:fem_rep_global}  is multiplied by the corresponding mode, $\{ U^k\}$:
\begin{equation}
\{ U^k\}^T\{ \Sigma \} = \{ U^k \}^T\sum_k \beta^k\{ U^k \}
\end{equation}
Enforcing the orthogonality relation isolates $\beta^k$:
\begin{equation}
\beta^k  = \{ U^k\}^T\{ \Sigma \}  
\end{equation}
Note, this is not the case of interest here as we are attempting to reconstruct a distribution from data that lack information regarding the spatial variations.  However, the determination of coefficients from a known distribution will be used later to assess the recovery method results.

\section{Numerical Methodology}
\label{sec:fe_methodology}

The recovery of intra-grain spatial variations in the stress distributions is accomplished by determining the weights for the harmonic expansions within all of the grains by minimzing an equilibrium-based objective function.
As the harmonic functions are defined with respect to a finite element mesh, the finite element discretization is intimately linked to the harmonic functions and thus the objective function. 
This dependency is implicit in the matrix development that follows.\\

\leftline{\it Grain boundary traction objective function}

The stress vector can be evaluated at any point, $\bf{x}_0$, within the grain domain using the harmonic expansion as:
\begin{equation}
 \Sigma_i({\bf{x}}_0)= \sum_k \beta^k_i u^k({\bf{x}}_0)   = \sum_k\beta^k_i u^k_0  
\label{eq:S_point}
\end{equation}
Likewise, the traction vector can be expressed at $\bf{x}_0$ 
\begin{equation}
\{ t \} = [ \hat n ] \{\Sigma\}  =  [ \hat n ] \sum_k\beta^\alpha_i u^k_0 =  [ \hat n ]  \sum_k \beta^k_i [ N({\bf{x}}_0)] \{ U\} ^k
\label{eq:matrix_cauchy}
\end{equation}
To construct a matrix expression of the objective function in terms of unknown expansion weights, we introduce $[ \hat m ] $ as
\begin{equation}
[ \hat m ]  = \left[ 
\begin{array}{c c c c c c}
                [u^k_0] &  0 &  0 &  0 &  0 &  0 \\
                0 &   [u^k_0]&  0 &  0 &  0 &  0 \\
                0 &  0 &   [u^k_0]&  0 & 0  &  0 \\
                0 &  0 &  0 &  [u^k_0]&  0 &  0 \\
                0 &  0 &  0 &  0 &   [u^k_0] &  0 \\
                0 &  0 &  0 &  0 &  0 &  [u^k_0]
 \end{array}
                \right]
\label{eq:matrix_U}
\end{equation}
where
\begin{equation}
 [u^k_0]  = \left[ 
\begin{array}{ccccc}
                [u^1({\bf{x}}_0)] &  [u^2({\bf{x}}_0)]  &  [u^2({\bf{x}}_0)]  & ... &  [u^{n^m}({\bf{x}}_0)]  
 \end{array}
                \right]
\label{eq:matrix_U_i}
\end{equation}
The traction vector can then be expressed in terms of the known normal vector and the full 
set of expansion weights  for all six components of stress as:
\begin{equation}
\{ t \} = [ \hat n ] [\hat m ] \{ \beta \} 
\label{eq:matrix_cauchy_beta}
\end{equation}
where
\begin{equation}
[ \hat n ]  = \left[ 
\begin{array}{c c c c c c}
                n_x &  0 &  0 &  0 &  n_z &  n_y \\
                0 &  n_y &  0 &  n_z &  0 &  n_x\\
                0 &  0 &  n_z &  n_y &  n_x &  0
 \end{array}
                \right]
\label{eq:matrix_normal}
\end{equation}
and
\begin{equation}
\{ \beta \} = \left\{
\begin{array}{c }
                \{\beta^k\}_1 \\
                \{\beta^k\}_2 \\
                \{\beta^k\}_3 \\
                \{\beta^k\}_4 \\
                \{\beta^k\}_5 \\
                \{\beta^k\}_6

 \end{array}
                \right\}
\label{eq:matrix_beta}
\end{equation}
Here the subscripts (numbered 1 to $n^m$) refer to the stress component (as per Equation~\ref{eq:S_sixvector}), while the superscript $k$ refers to the mode number:
\begin{equation}
 \{\beta^k\}_i= \left\{
\begin{array}{c }
                \beta^1 \\
                \beta^2 \\
                \beta^3 \\
                 ... \\
                \beta^{n^m}

 \end{array}
                \right\}
\label{eq:matrix_beta_i}
\end{equation}

The equilibrium violation vector for at a grain interface,  as defined in Equation~\ref{eq:boundary_resultant}, becomes: 
\begin{equation}
\vctr{f}^b = \vctr{t}^i + \vctr{t}^j  = [ \hat n]^i [\hat m ]^i \{ \beta \}^j +  [ \hat n]^j [\hat m ]^j \{ \beta \}^j  = 
\left[    
\begin{array}{cc}
              [\tilde m]^i& 0  \\
           0 & [\tilde m]^j             
              \end{array} 
               \right] 
              \left\{\begin{array}{c }
                 \{\beta^k\}^i\\
                 \{\beta^k\}^j
              \end{array}\right\}
                \label{eq:boundary_resultant_matrix}
\end{equation}
and the corresponding contribution to the objective function is:
\begin{equation}
\left( F^b \right)^{ij}= 
              \left\{\begin{array}{c }
                 \{\beta^k\}^i\\
                 \{\beta^k\}^j
              \end{array}\right\}^T
              \left[    
\begin{array}{cc}
              [\tilde m]^i& 0  \\
           0 & [\tilde m]^j             
              \end{array} 
               \right]^T 
\left[    
\begin{array}{cc}
              [\tilde m]^i& 0  \\
           0 & [\tilde m]^j             
              \end{array} 
               \right] 
              \left\{\begin{array}{c }
                 \{\beta^k\}^i\\
                 \{\beta^k\}^j
              \end{array}\right\}
              =
                            \left\{\begin{array}{c }
                 \{\beta^k\}^i\\
                 \{\beta^k\}^j
              \end{array}\right\}^T
              \left[    
                  \tilde M^{ij}
               \right] 
              \left\{\begin{array}{c }
                 \{\beta^k\}^i\\
                 \{\beta^k\}^j
              \end{array}\right\}
                \label{eq:boundary_residual_matrix}
\end{equation}
Summing over all the grain boundary interface points gives:
\begin{equation}
F^b = \sum_{ij_{\rm pairs}} \left( F^b \right)^{ij}
\label{eq:boundary_residual_total_matrix}
\end{equation}
In the present implementation, the number of point is designated as  ${ij_{\rm pairs}}$ and
consists of the centroids of all surface elements lying on the grain boundaries. 
Free surface elements are a special case of this in which the pair is degenerate and has only one element of the pair.  This is the desired result for free surfaces  -- namely, that the traction on free surface is zero.\\

\leftline{\it Local equilibrium objective function}

Within grain interiors, the equilibrium violation vector is:
\begin{equation}
 \{ f^v\}^{qp}=  
  \left\{
\begin{array}{c }
                \frac{\partial \sigma_{xx}}{\partial x} + \frac{\partial \sigma_{xy}}{\partial y} + \frac{\partial \sigma_{xz}}{\partial z} \\
                 \frac{\partial \sigma_{xy}}{\partial x} + \frac{\partial \sigma_{yy}}{\partial y} + \frac{\partial \sigma_{yz}}{\partial z} \\
                \frac{\partial \sigma_{xz}}{\partial x} + \frac{\partial \sigma_{yz}}{\partial y} + \frac{\partial \sigma_{zz}}{\partial z} 
 \end{array}
                \right\}^{qp}
\label{eq:divsigma_force}
\end{equation}
The superscript, $qp$ refers to a quadrature point.  In the present implementation, this vector is evaluated at all of the  quadrature points  over all of the elements.  
We note that the stress gradient appearing in Equation~\ref{eq:divsigma_force} can be expressed in terms of the spatial derivatives of the harmonic functions.
For the $x$-dircection, for instance, the result is: 
\begin{equation}
\frac{\partial \Sigma_1}{\partial x} = \sum_k \beta^k_1 \frac{\partial u^k({\bf{x}}) }{\partial x}
= \left[  \frac{\partial u^k({\bf{x}}) }{\partial x}\right]  \{\beta^k\}_1
\label{eq:stress_grads}
\end{equation}
To obtain the matrix form of the objective function, we first define $ \left[ \bar m \right]$ as
\begin{equation}
\left[ \bar m \right] =  \left[ 
\begin{array}{c c c c c c}
                \left[  \frac{\partial u^k}{\partial x}\right]  &  0 &  0 &  0 &  \left[  \frac{\partial u^k }{\partial x}\right]  & \left[  \frac{\partial u^k}{\partial y}\right]  \\
                0 &  \left[  \frac{\partial u^k}{\partial y}\right]  &  0 &  \left[  \frac{\partial u^k}{\partial z}\right]  &  0 &  \left[  \frac{\partial u^k }{\partial x}\right] \\
                0 &  0 &  \left[  \frac{\partial u^k}{\partial z}\right]  &  \left[  \frac{\partial u^k}{\partial y}\right]  &  \left[  \frac{\partial u^k}{\partial x}\right]  &  0
 \end{array}
                \right] 
\label{eq:mbar}
\end{equation}
and with it write the equilibrium violation vector for local equilibrium as 
\begin{equation}
 \{ f^v\}^{qp}  =   \left[ \bar m \right] \{ \beta \} 
\label{eq:divsigma_force_2}
\end{equation}
With these matrices in hand, the objective function is written in term of the equilibrium violation vector for one point:
    \begin{equation}
   ( F^v)^{qp} =    \{ \beta \}^T    \left[ \bar m \right]^T  \left[ \bar m \right] \{ \beta \}     =   \{ \beta \}^T \left[ \bar M^{qp} \right] \{ \beta \} 
\label{eq:divsigma_force_3}
\end{equation}
and then summed over all the quadrature points:
\begin{equation}
F^v = \sum_{{\rm qp}} \left( F^v \right)^{qp}
\label{eq:local_equil_total_matrix}
\end{equation}

Note that the summation processes  of Equations~\ref{eq:boundary_residual_total_matrix} and \ref{eq:local_equil_total_matrix} consist of assembling local objective function matrices into global matrices that include the harmonic expansion modes for all the grains and all the stress components.   These matrices are square with dimension of 
$n^g \times n^m \times n^c$,
where $n^g$ is the number of grains.  However, they are sparse, an attribute that can be exploited within \matlab.
For illustration, a 1000 grain polycrystal with a 10 mode expansion would require 60,000 degrees of freedom (1000x10x6) for the spatial representation of stress.   For a mesh for a 1000 grain polycrystal with  500,000 nodes, the representation of a stress distribution with intragrain continuity the degrees of freedom would number 3,000,000.  For this case, the reduced-order representation using the expansion reduces the degrees of freedom count by a factor of 50.

\section{Recovery of the Stress Distribution in a FCC Polycrsytal}
\label{sec:demo_application}
In \cite{Mechmonics_2021}, three variants of a polycrystal sample were examined.  For all three,  harmonic modes were computed and  correlations of the modes with the grain axes were examined.  
In addition, the deformations of the polycystals for a nominal extensional strain of $1\%$ were
simulated with \fepx.  Using the stress distributions from these simulations,   the evolutions of the expansion weights for the stress were evaluated through the elastic-plastic transition.
To illustrate the stress recovery methodology here, we examine synthetic diffraction data for one of the three sample variants -- the sample having grains with the most uniform sizes and most spherical shapes (designated as HUHS).
The synthetic data was developed from the stress distribution computed using \mechmet\, by averaging the computed intra-grain distributions over the respective grains.  
This produced grain-averaged fields that mimic what might be obtained from experiments.
However, the full intra-grain distributions are known, so comparison of the reconstructed distributions to the original distributions is possible.

\subsection{Sample grain attributes and mechanical metrics}
The three variants all define a 100-grain sample and were generated with \neper\ . 
The individual tessellations for the three samples were created with different grain diameter distributions and different grain sphericity distributions by altering the seed attributes (as accommodated in the \neper\, input data).  
Here we use the one dsignated as HUHS for its high grain size uniformity and high sphericity (relative to the other samples).
  Mean and standard deviation for the size and sphericity are given in Table~\ref{tab:sample_stats}.
\begin{table}[ht]
\small
\centering
\begin{tabular}{| c | c |}
\hline
 Volume Mean &  0.020 	\\
 Volume SD&  0.0093 	\\
 Sphericity Mean & 0.862 	\\
 Sphericity SD  &  0.0287 	\\
 \hline
\end{tabular}
\caption{Diameter and sphericity statistical metrics for the HUHS sample. }
\label{tab:sample_stats}
\end{table}
 
\begin{figure}[htbp]
	\centering
	\begin{subfigure}{.3\textwidth}
		\centering
		\includegraphics[width=1\linewidth]{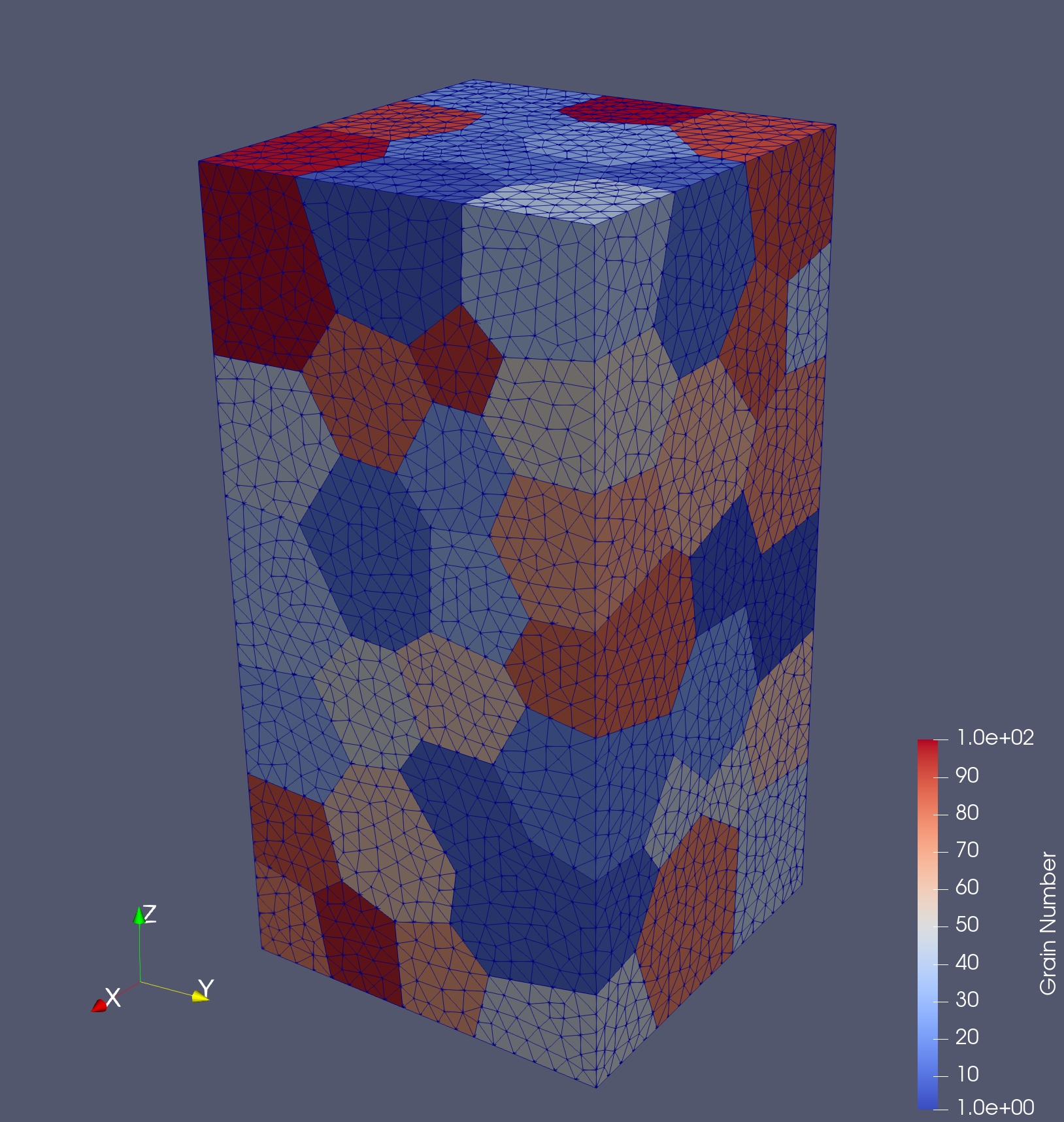}
		\caption{ }
		\label{fig:dia0p15_sph0p03_mesh}
	\end{subfigure}%
	\quad
		\begin{subfigure}{.3\textwidth}
		\centering
		\includegraphics[width=1\linewidth]{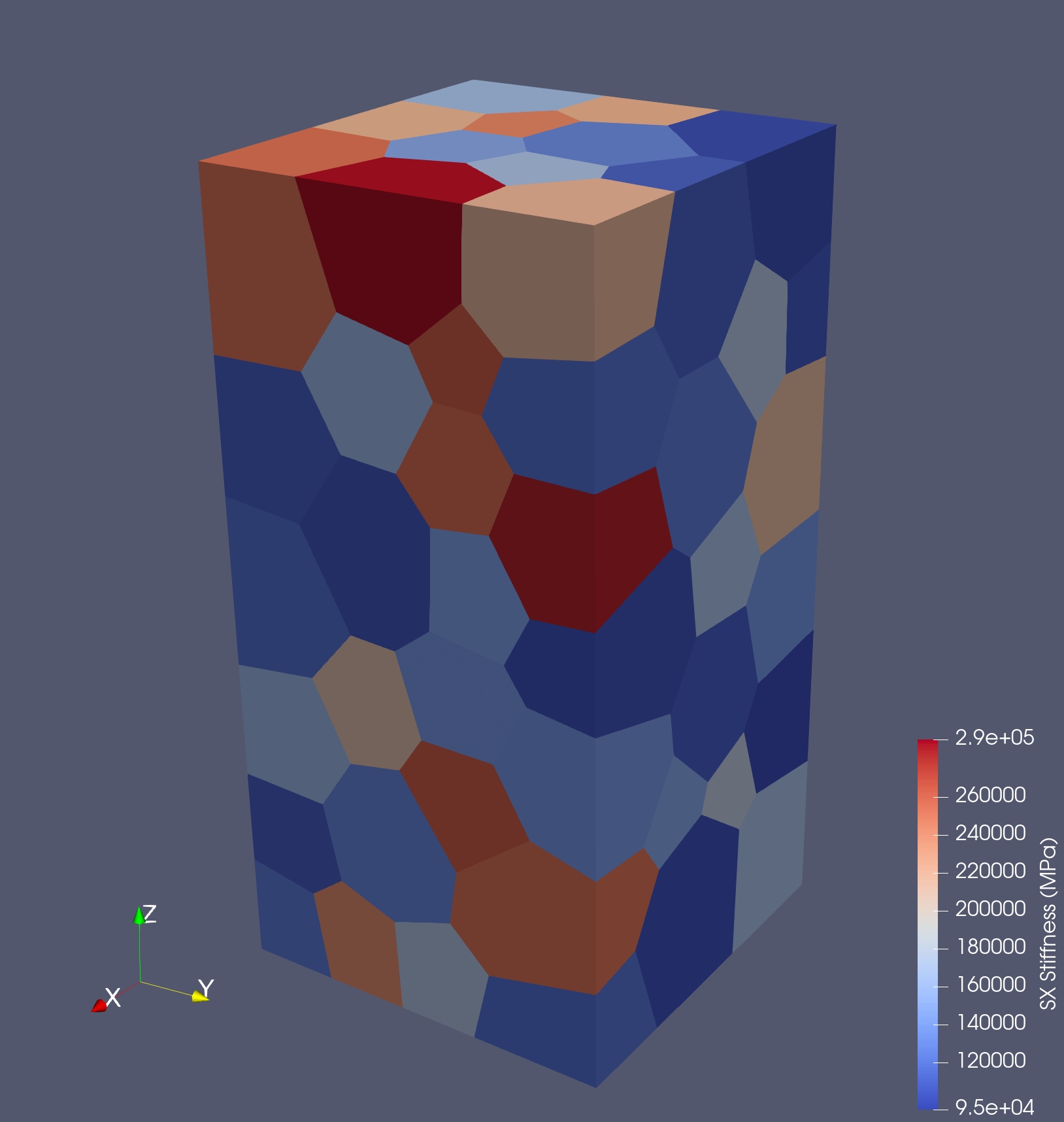}
		\caption{ }
		\label{fig:dia0p15_sph0p03_sxstiff}
	\end{subfigure}%
	\quad
	\begin{subfigure}{.3\textwidth}
		\centering
		\includegraphics[width=1\linewidth]{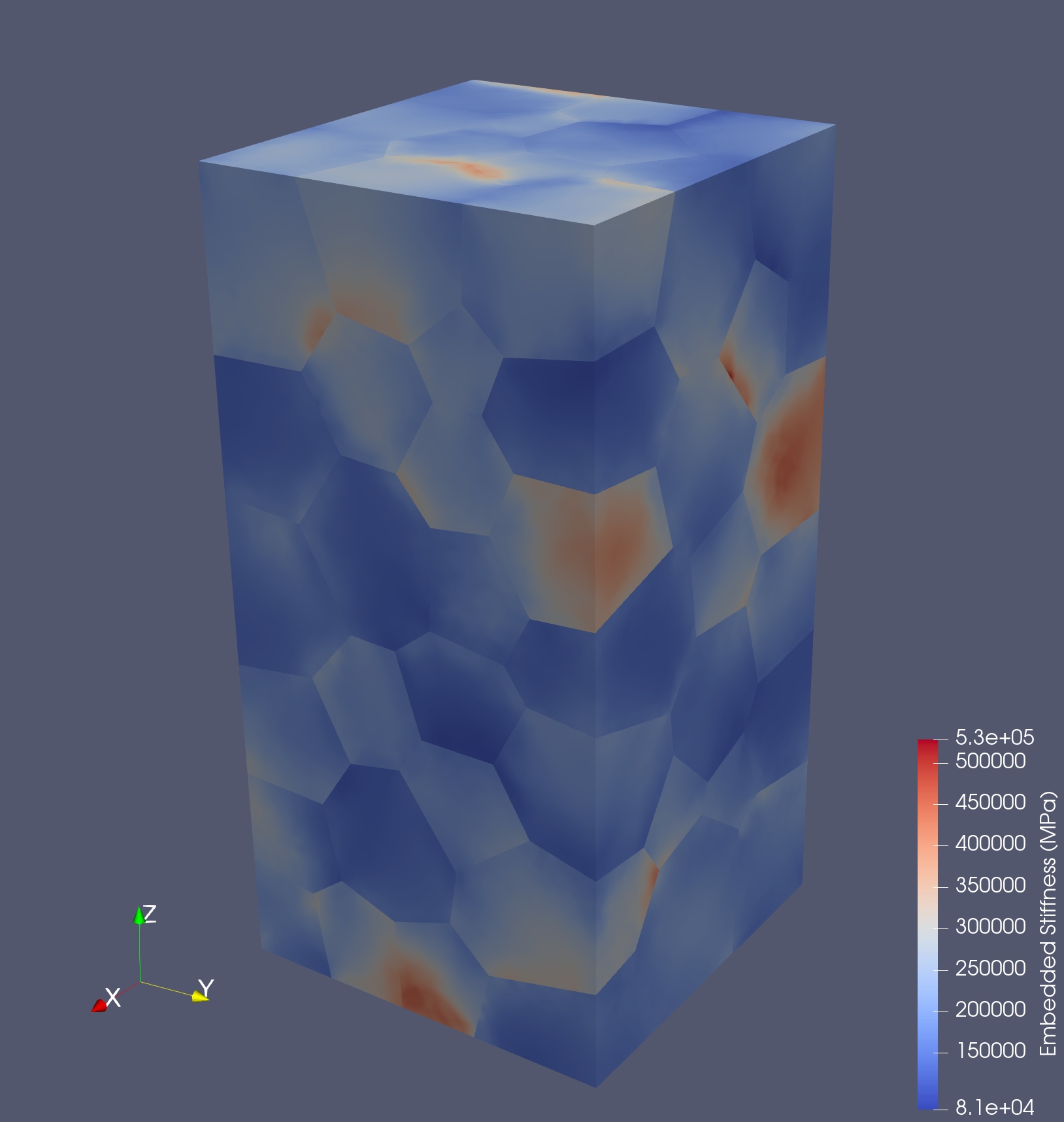}
		\caption{ }
		\label{fig:dia0p15_sph0p03_embstiff}
	\end{subfigure}%
	\caption{Attributes of the HUHS sample:  (a) mesh over grains; (b) single crystal stiffness; and (c) embedded crystal stiffness}
		\label{fig:sample_dia0p15_sph0p03}
\end{figure}

  Figure~\ref{fig:sample_dia0p15_sph0p03} shows several aspects of this microstructure.
  Part(a) shows the grains with the finite element mesh.  For all variants, the meshing controlled to produce virtual polycrystals with about 100K elements.  
  Lattice orientations were assigned to the grains randomly from a uniform distribution and are   
  uniform over the grains initially.  
  Two elastic properties are shown:  the z-direction (axial) directional elastic modulus (Part(b))  and the embedded stiffness for z-direction (axial) tension (Part(c)), both evaluated for AL6XN~\cite{pos_daw_mmta_2019a}.

\subsection{Harmonic modes}
\label{sec:harmonicmodes}
\begin{figure}[h!]
	\centering
	\begin{subfigure}{.3\textwidth}
		\centering
		\includegraphics[width=1\linewidth]{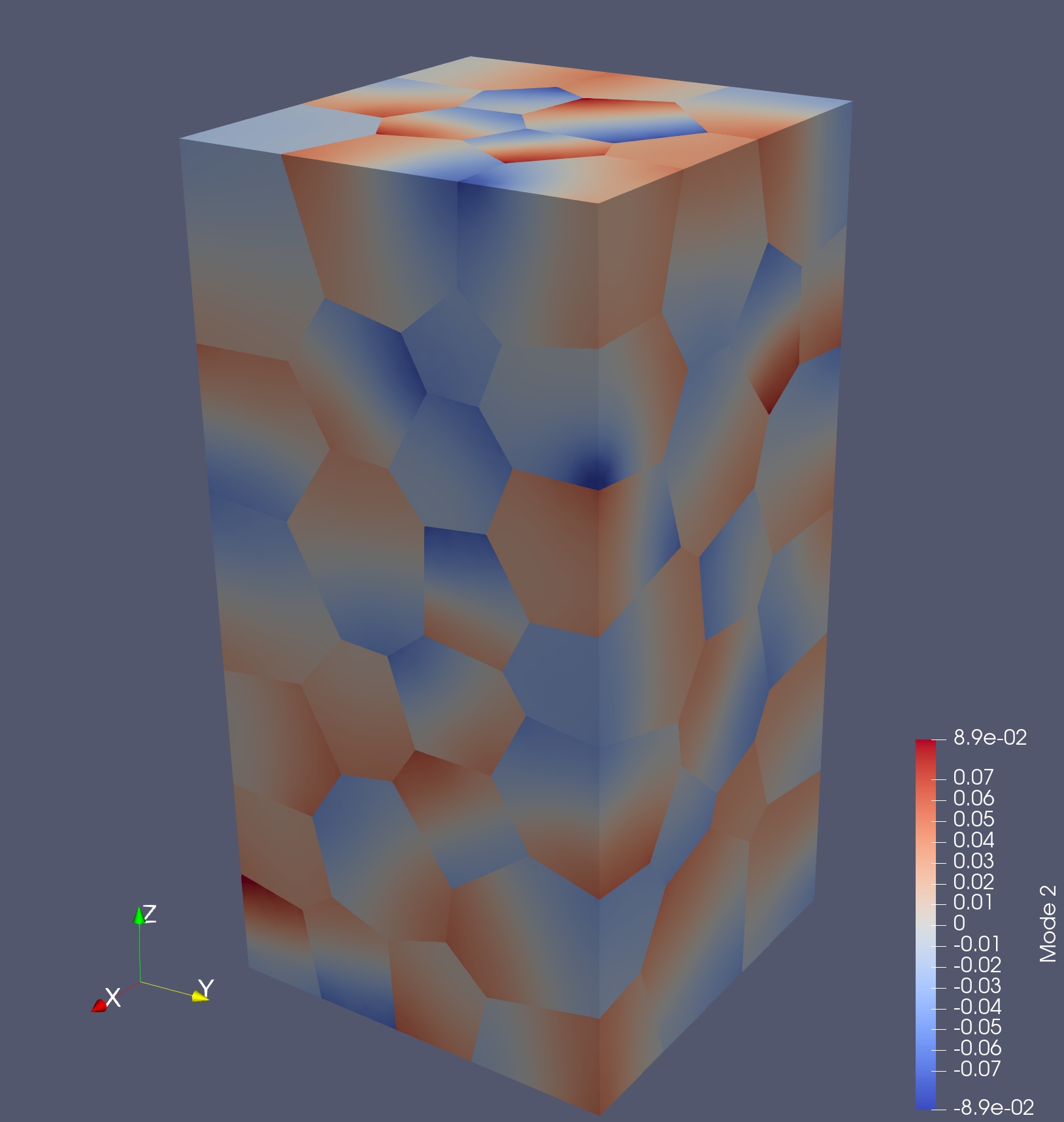}
		\caption{ }
		\label{fig:dia0p15_sph0p03_mode1}
	\end{subfigure}%
	\quad
	\begin{subfigure}{.3\textwidth}
		\centering
		\includegraphics[width=1\linewidth]{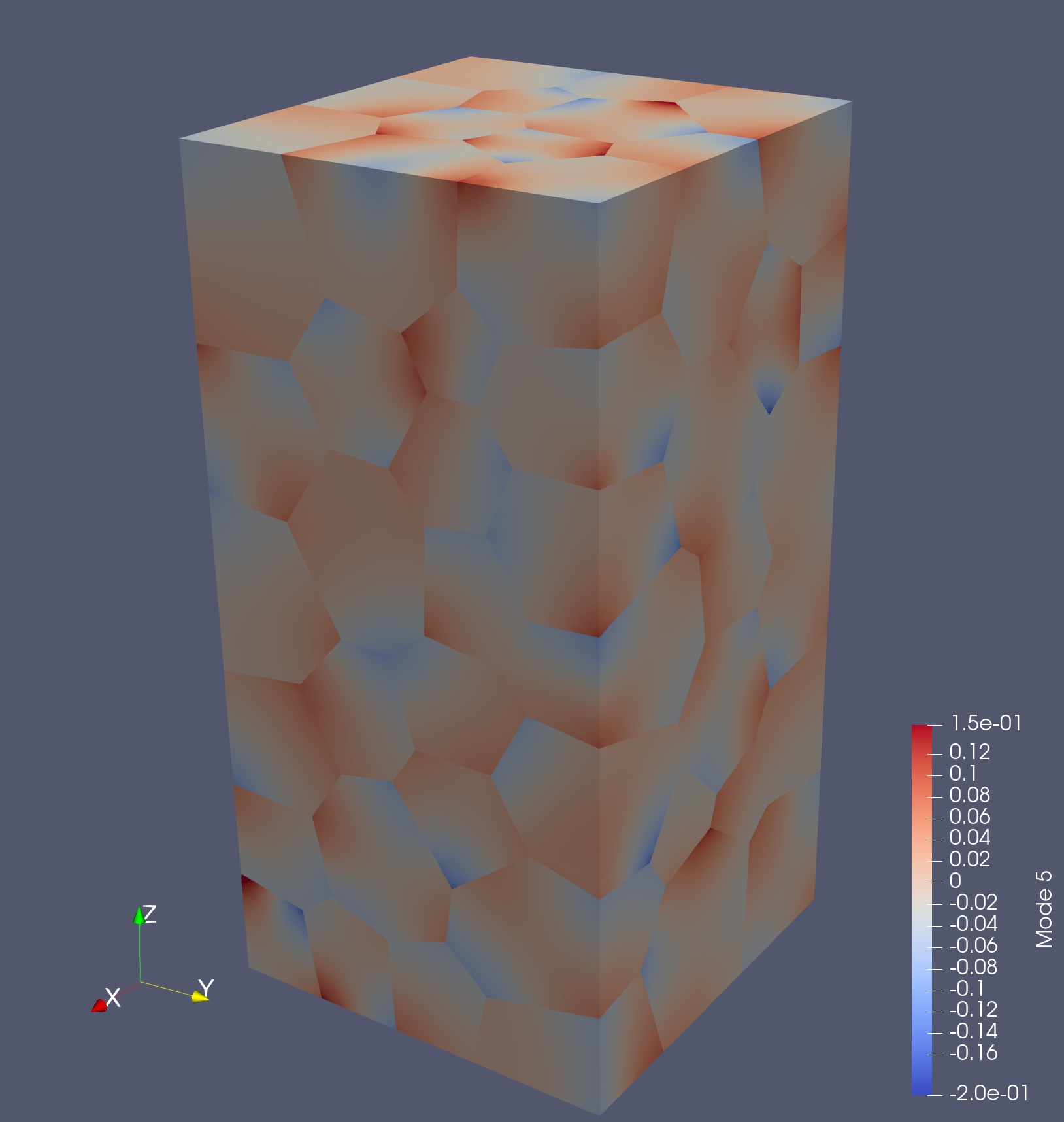}
		\caption{ }
		\label{fig:dia0p15_sph0p03_mode4}
	\end{subfigure}%
        \quad
	\begin{subfigure}{.3\textwidth}
		\centering
		\includegraphics[width=1\linewidth]{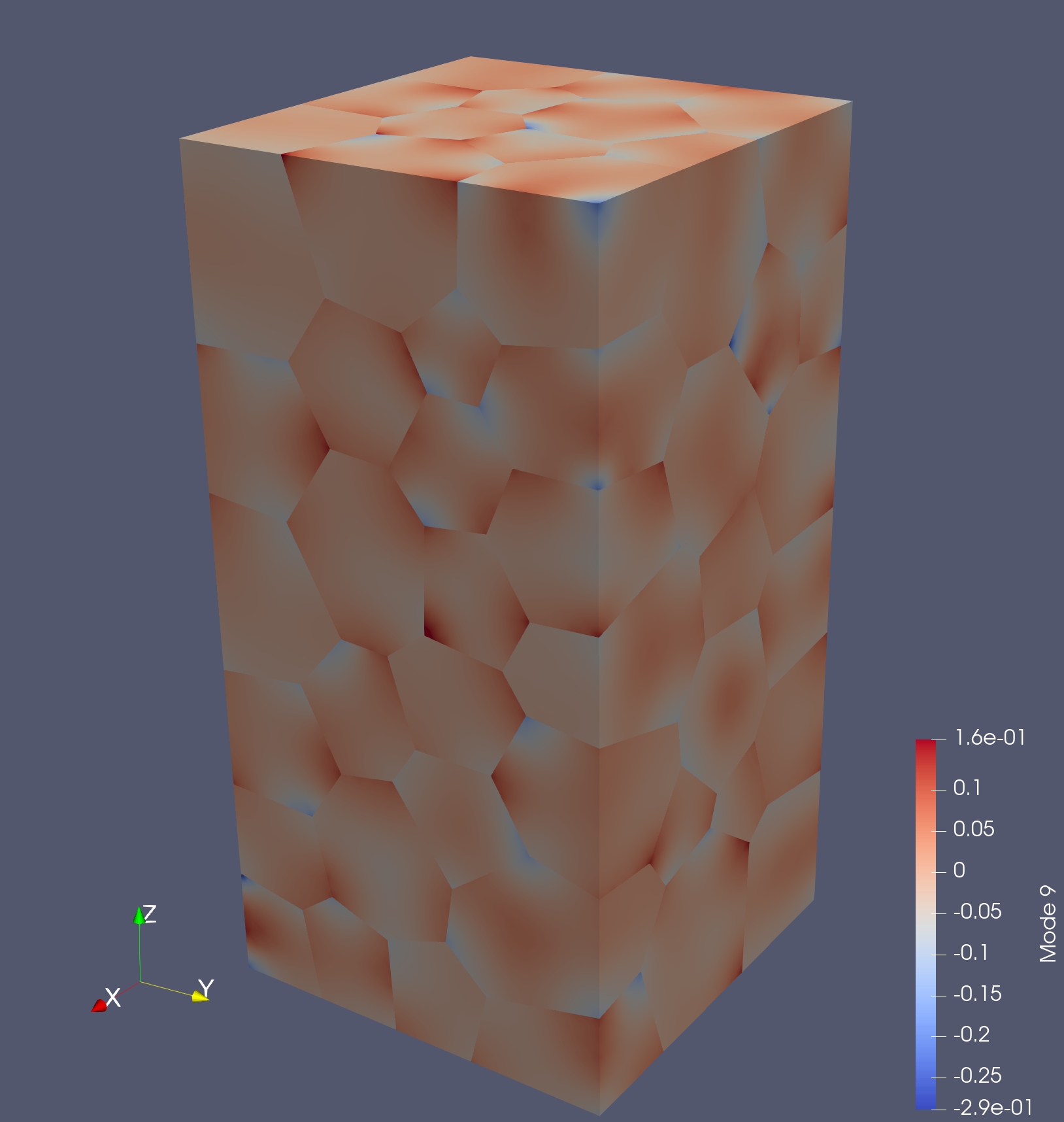}
		\caption{ }
		\label{fig:dia0p15_sph0p03_mode8}
	\end{subfigure}%
	\caption{Harmonic modes of the HUHS sample:  (a) Mode 2; (b) Mode 5; and (c) Mode 9.}
		\label{fig:harmonicmodes_dia0p15_sph0p03}
\end{figure}

Discrete harmonic modes were determined with \mechmonics\, according to the approach laid out in \cite{Mechmonics_2021}.
For the demonstration objective here,  the first 28 modes are discussed.   
Modes 2, 5 and 9 are shown for each sample in Figure~\ref{fig:harmonicmodes_dia0p15_sph0p03}.
Mode 1  (not shown) is the constant mode for each set of harmonic modes. 
Modes other than the first are not constant, but rather vary over the grain with increased complexity as the mode number increases.
Mode 2 has monotonically increasing or decreasing distributions. 
However, its gradients are not constants (so the mode is not linear).
Generally speaking, the values of the gradient of Mode 2 are lower than corresponding values for the other modes.
Modes 5 and 9 have more complex distributions and definitely are not monotonic over grains.
This trend continues as the mode number increases.  
 Note that in general the mode patterns do not  repeat the same pattern from grain to grain.

\begin{figure}[h!]
	\centering
	\begin{subfigure}{.3\textwidth}
		\centering
		\includegraphics[width=1\linewidth]{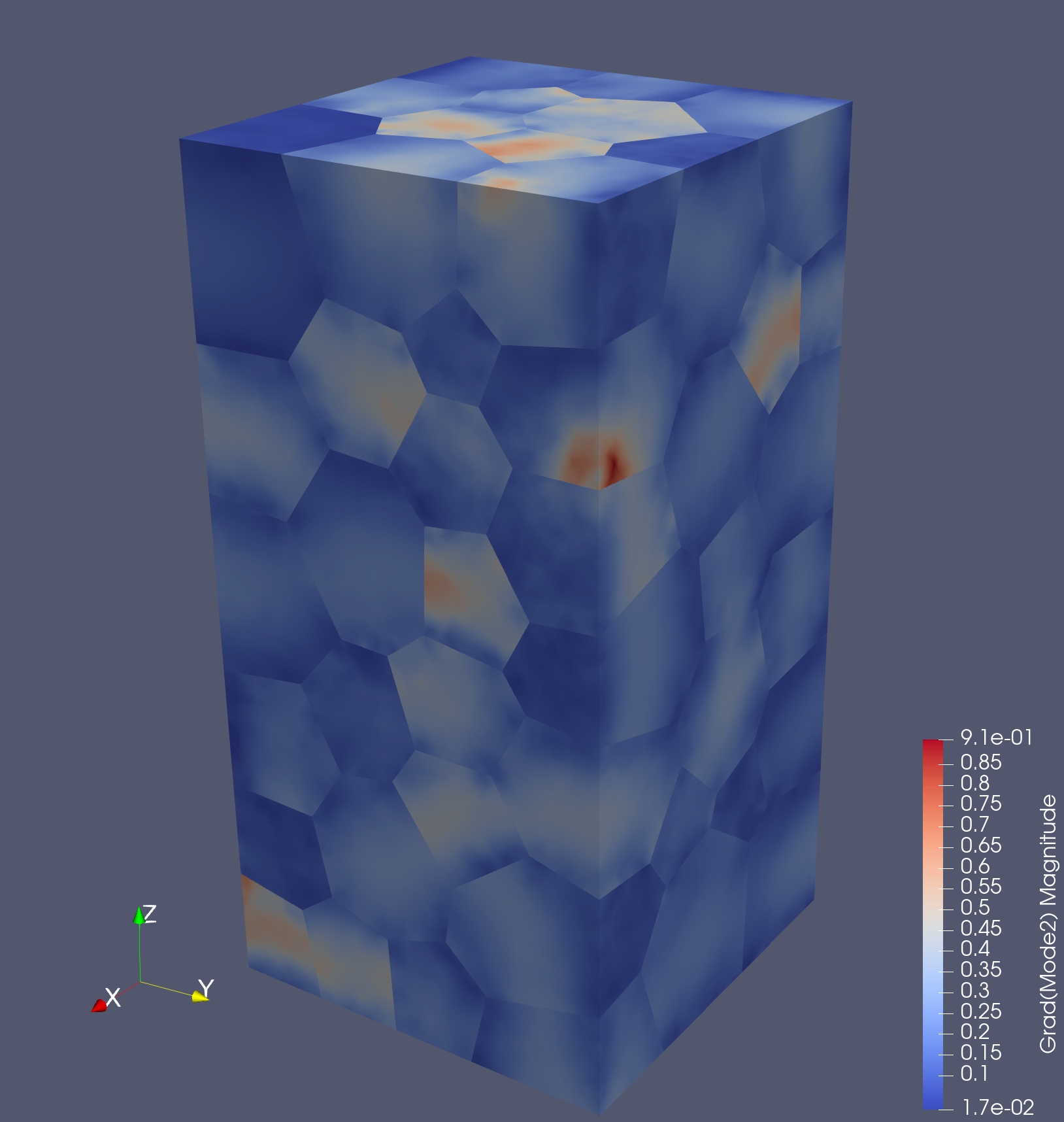}
		\caption{ }
		\label{fig:dia0p15_sph0p03_gradmode1}
	\end{subfigure}%
		\quad
	\begin{subfigure}{.3\textwidth}
		\centering
		\includegraphics[width=1\linewidth]{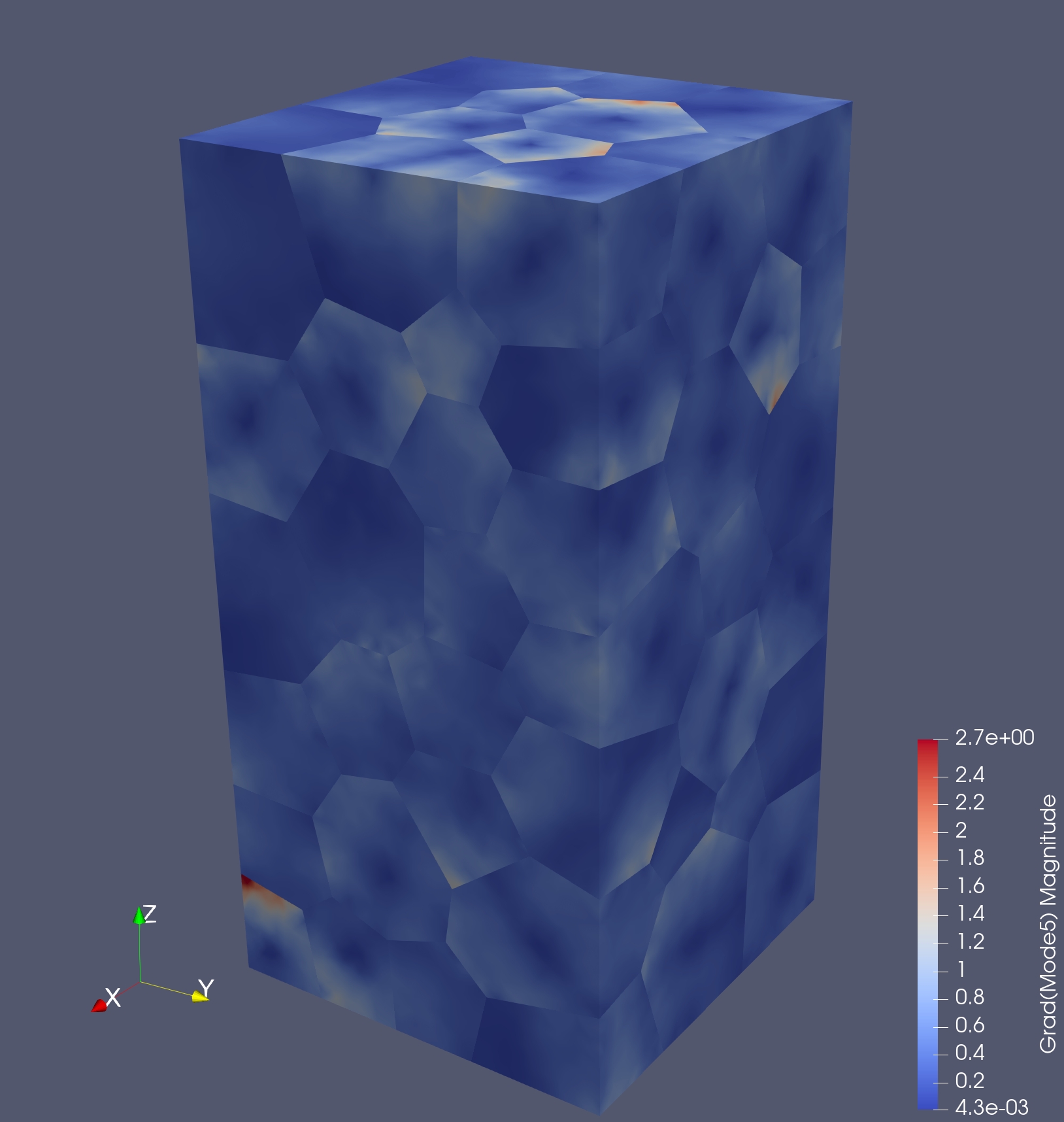}
		\caption{ }
		\label{fig:dia0p15_sph0p03_gradmode4}
	\end{subfigure}%
	\quad
	\begin{subfigure}{.3\textwidth}
		\centering
		\includegraphics[width=1\linewidth]{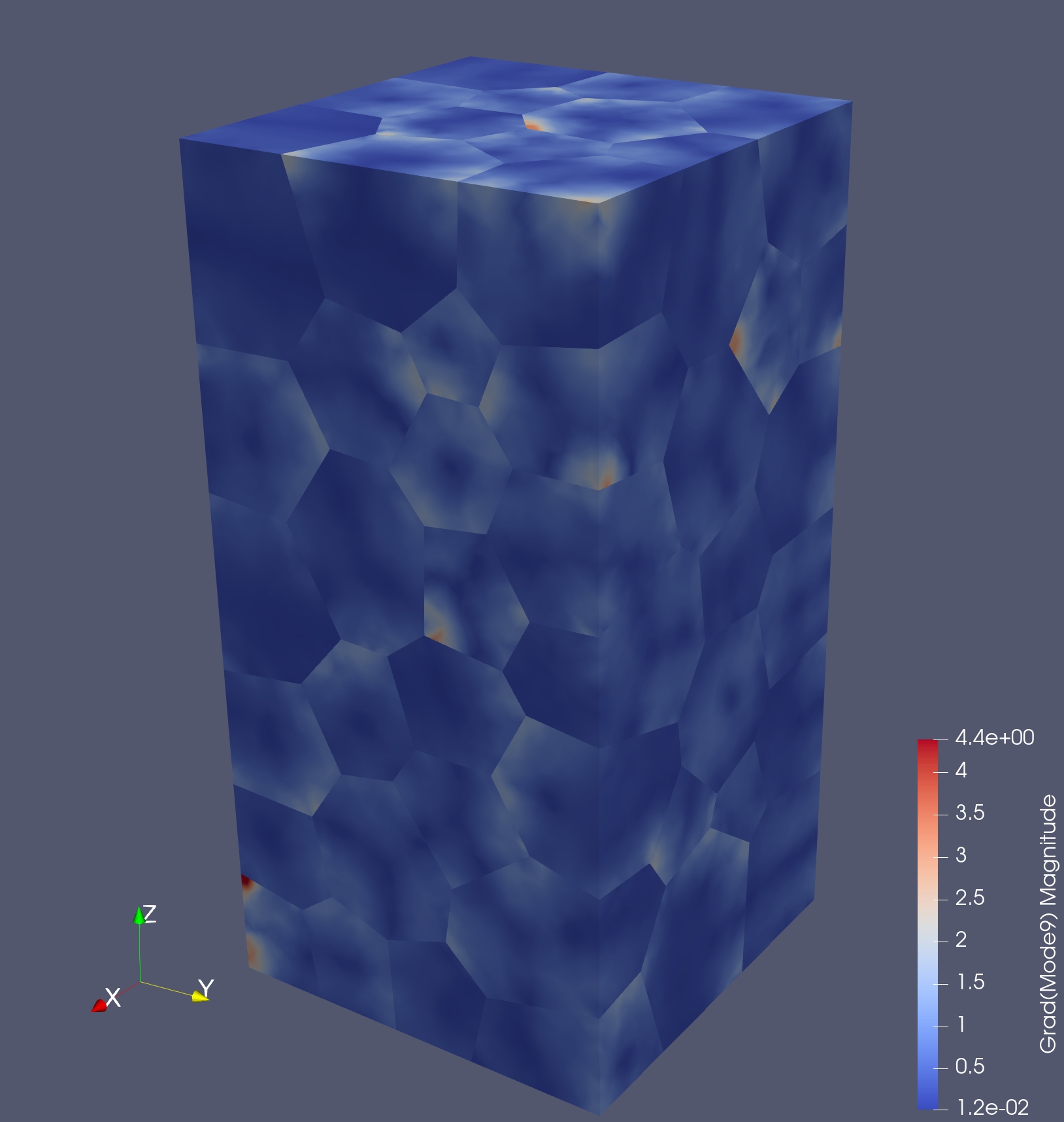}
		\caption{ }
		\label{fig:dia0p15_sph0p03_gradmode8}
	\end{subfigure}%
	\caption{Magnitudes of the harmonic mode gradient vectors for the HUHS sample:  (a) Mode 2; (b) Mode 5; and (c) Mode 9.}
		\label{fig:harmonicmodesgrad_dia0p15_sph0p03}
\end{figure}

Figure~\ref{fig:harmonicmodesgrad_dia0p15_sph0p03}
shows the magnitudes of the vector gradients for Modes 2, 5 and  9 for the HUHS sample.
Although the direction of the modes (the gradient vector) appear to be haphazard, there is a correlation with the grain axes, as discussed in more detail in \cite{Mechmonics_2021}.

\subsection{Simulated stress distribution for axial extension loading}
\label{sec:mechmet_stress}
\begin{figure}[htbp]
  \centering
  \includegraphics[width=0.3\linewidth]{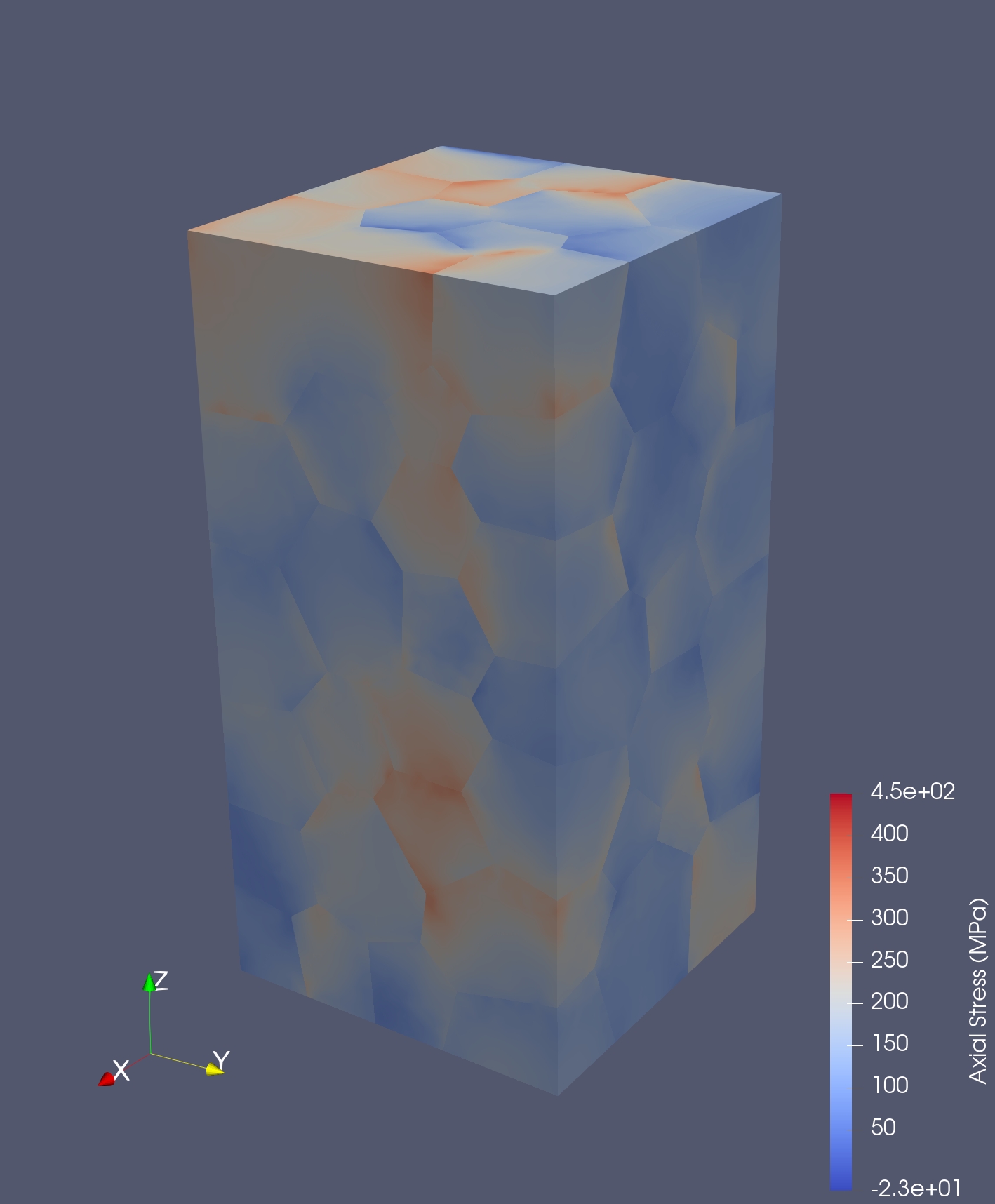}
  \caption{Axial stress under $z$-direction extension. }
  \label{fig:sample_axial_stress}
\end{figure}

A stress distribution for the sample was computed using \mechmet\, for a loading condition of $z-$direction extension to 0.1\% nominal strain.  
Based on elasticity theory,  \mechmet\, computes a displacement field and subsequently evaluates the strain field from kinematics and the stress field from the strain field and Hooke's law.  From the elemental quadrature  point stress tensors,  smoothed intra-grain distributions are evaluated at nodal points.  
The axial component of the stress is displayed in Figure~\ref{fig:sample_axial_stress}.

\begin{figure}[htbp]
	\centering
	\begin{subfigure}{.30\textwidth}
		\centering
		\includegraphics[width=1\linewidth]{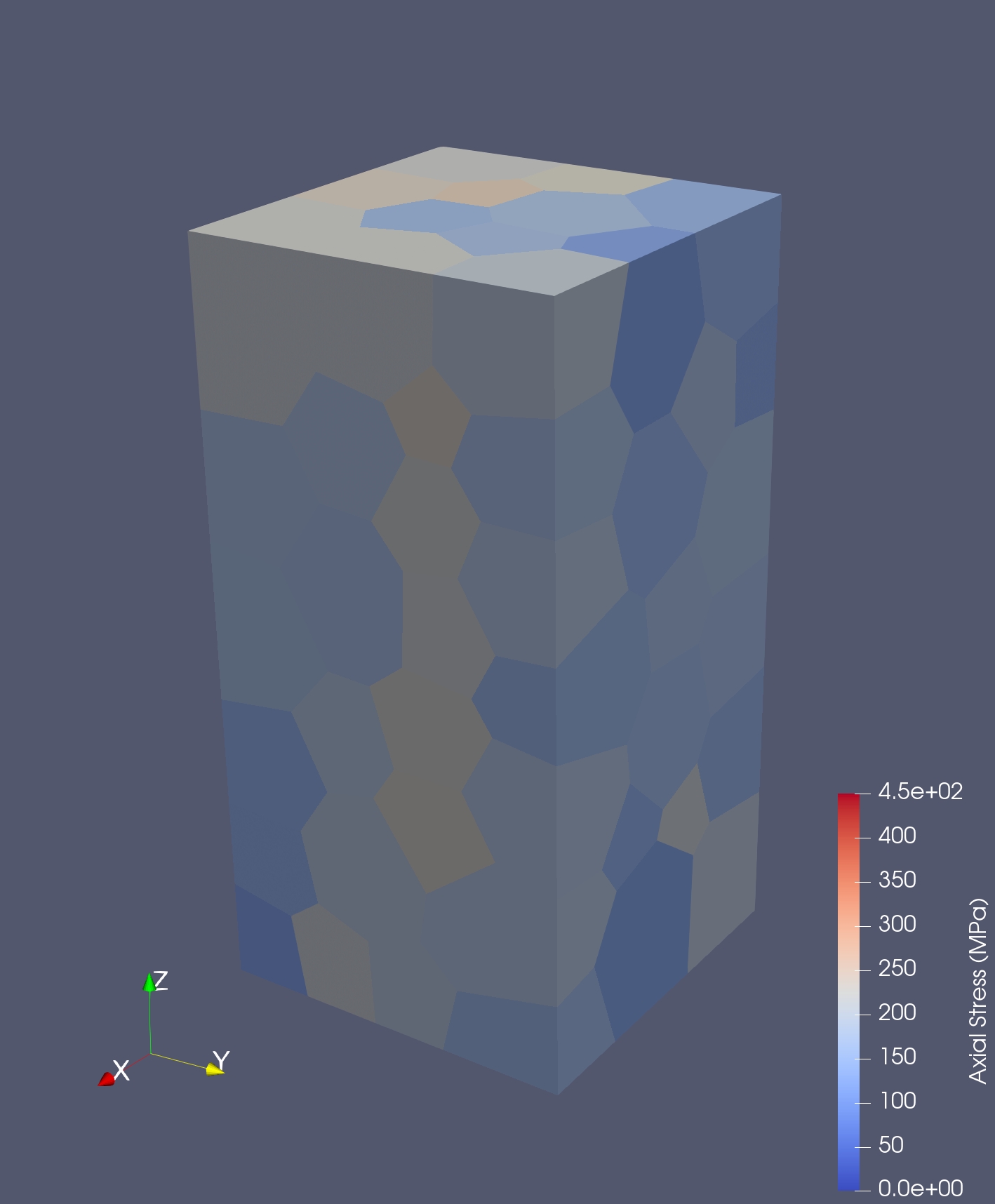}
		\caption{ }
		\label{fig:mechmet_1mode_sigzz}
	\end{subfigure}%
	\quad
	\begin{subfigure}{.30\textwidth}
		\centering
		\includegraphics[width=1\linewidth]{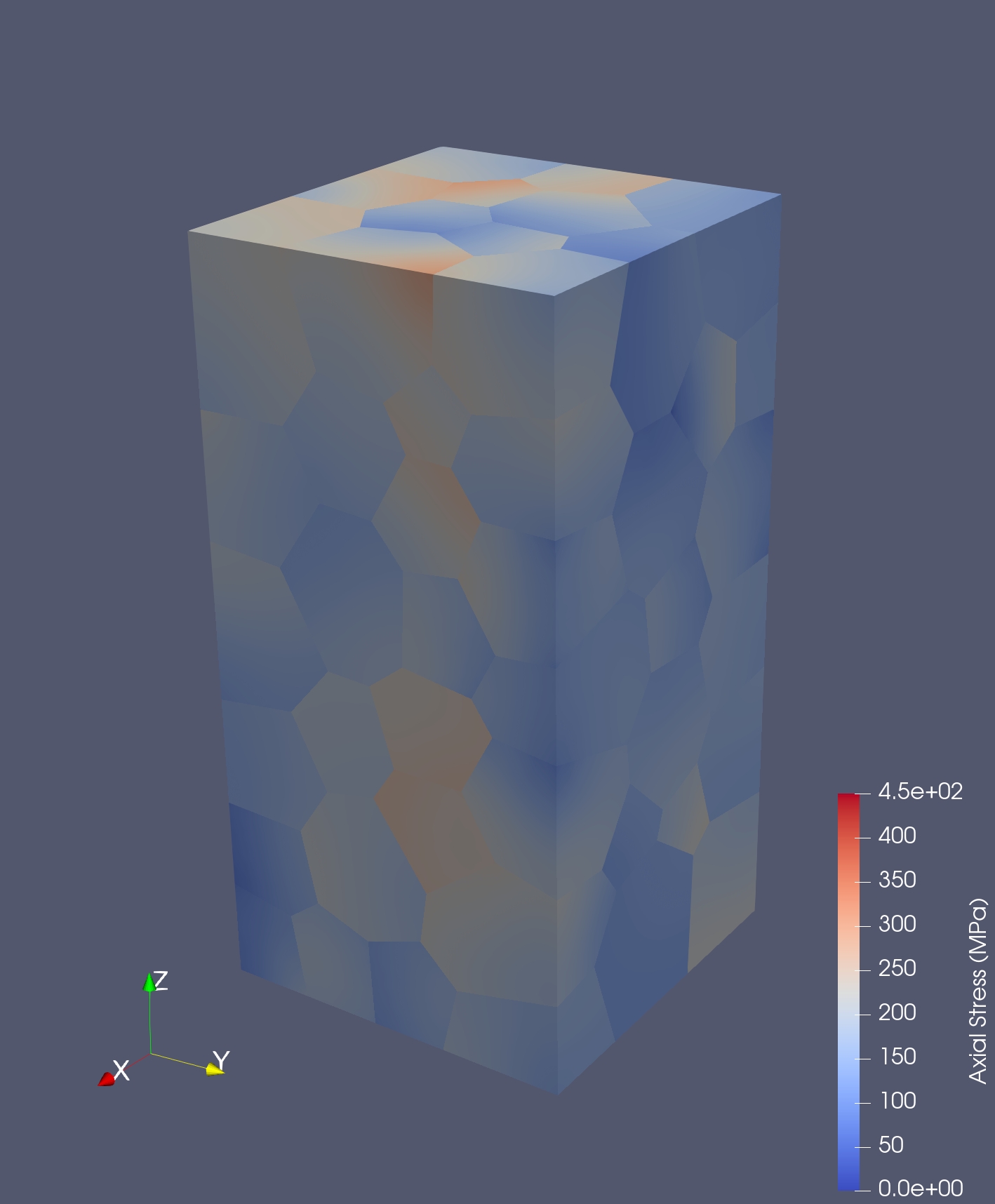}
		\caption{ }
		\label{fig:mechmet_4mode_sigzz}
	\end{subfigure}\\
	\begin{subfigure}{.30\textwidth}
		\centering
		\includegraphics[width=1\linewidth]{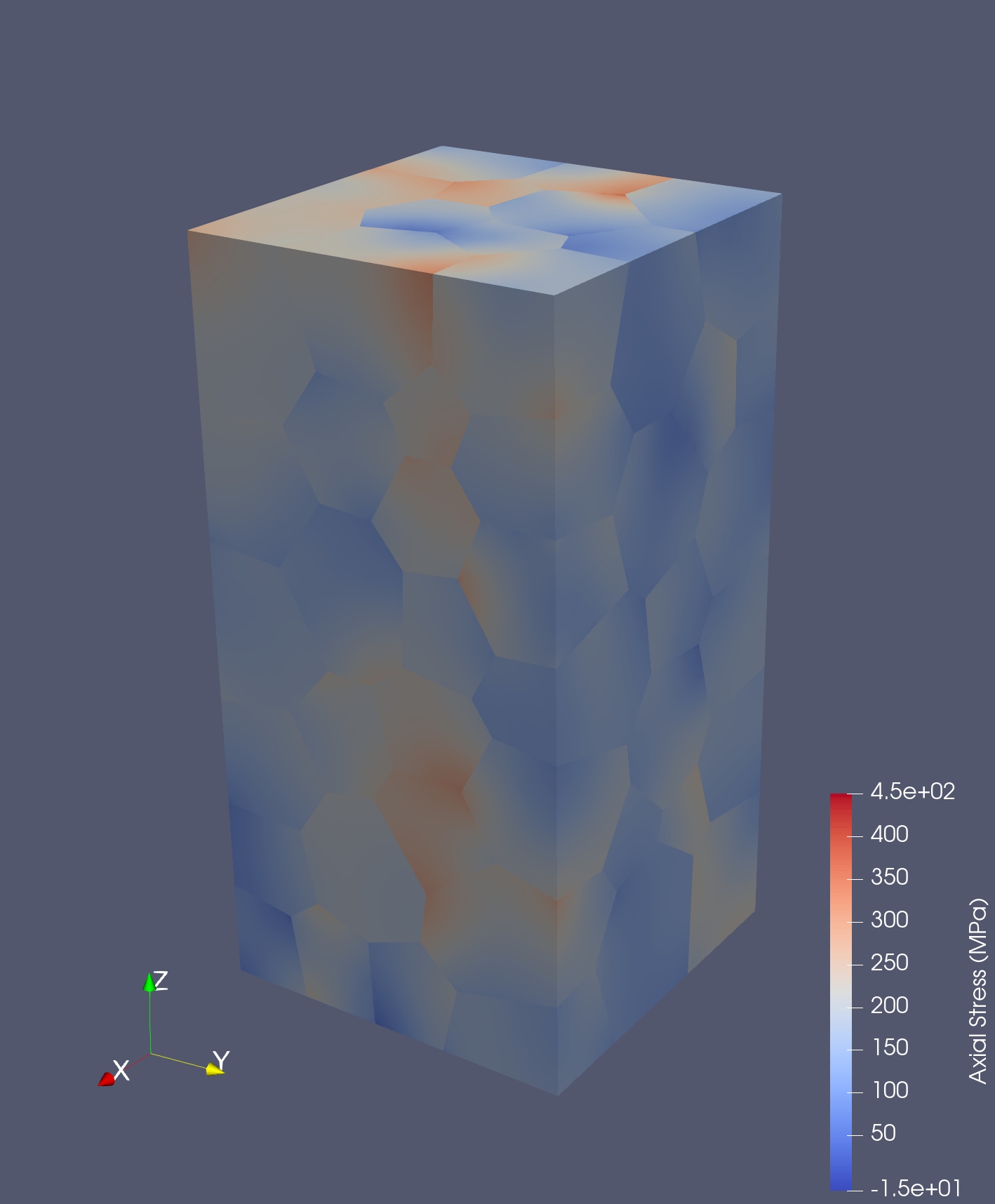}
		\caption{ }
		\label{fig:mechmet_10mode_sigzz}
	\end{subfigure}%
	\quad
	\begin{subfigure}{.30\textwidth}
		\centering
		\includegraphics[width=1\linewidth]{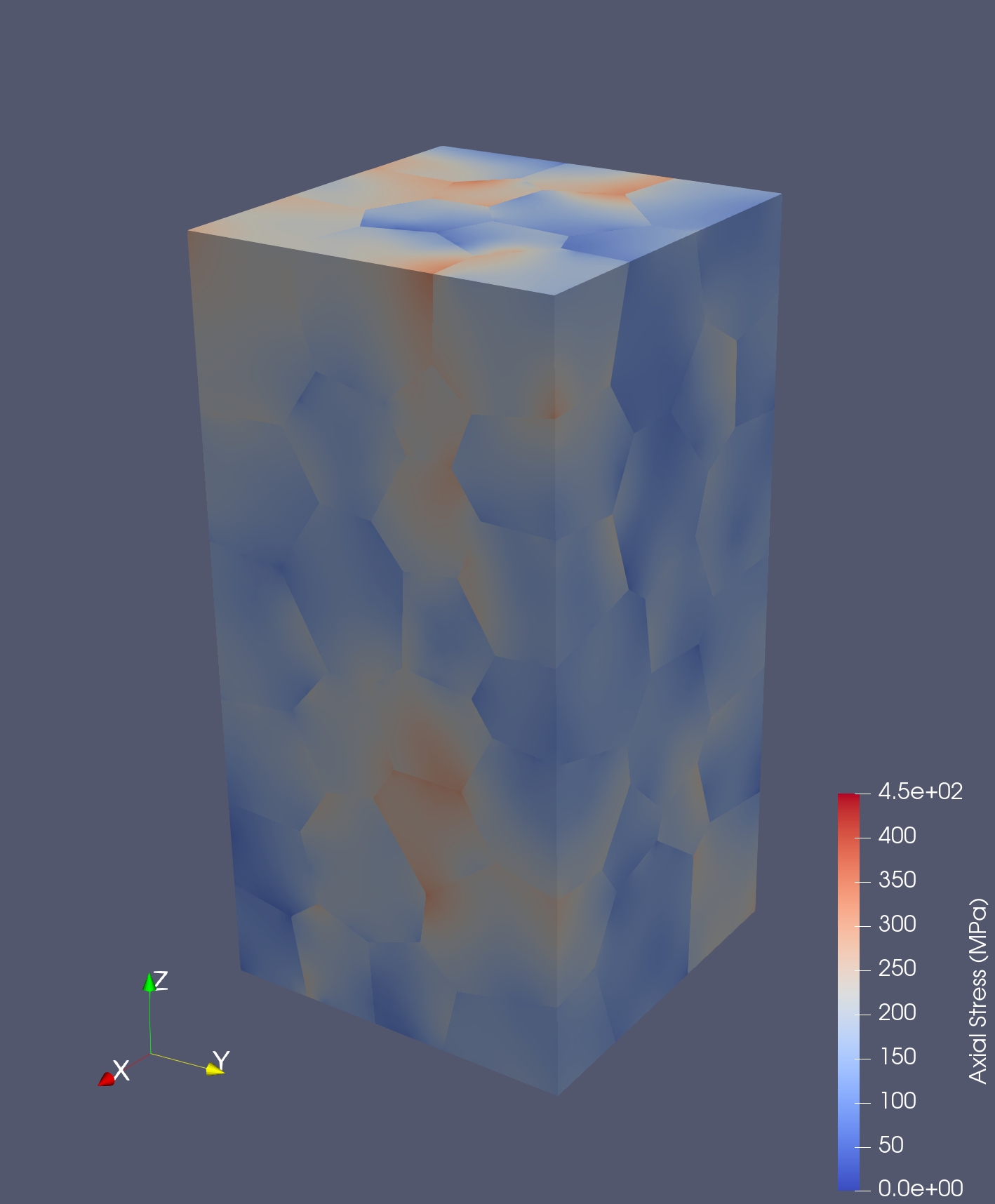}
		\caption{ }
		\label{fig:mechmet_28mode_sigzz}
	\end{subfigure}%
	\caption{Mode fits to the axial stress:  (a) 1 mode; (b) 4 modes; (c) 10 modes; and (d) 28 modes. }
		\label{fig:sample_axial_stress_modefits}
\end{figure}

Using this distribution, the weights  for 1, 4,  10 and 28 mode expansions were
computed for all 6 components of the stress
as outlined in Section~\ref{sec:harmonics_background}. 
The distributions for the axial stress component represented by these expansions are displayed in Figure~\ref{fig:sample_axial_stress_modefits}. 
Improvement of representation  with increasing numbers of modes is evident visually.
The expansions serve two uses here.
First, the grain-averaged stresses are known from the weights of the first mode of the expansion. These serve directly as the synthetic data.
Second, the distributions can be compared to the distributions obtained using the stress recovery procedure.  The comparisons can be made with the same number of modes for each, which facilitates isolating the influence of basing the objective function on equilibrium conditions.
\clearpage

\subsection{Reconstructed stress distribution using the methodology}
\label{sec:recovered_stress}

The grain-averaged stress computed with \mechmet\, stands in as synthetic HEDM data.  
The single-term expansion provides the averaged values, and has been shown for the axial component
of the stress in Figure~\ref{fig:mechmet_1mode_sigzz}.
With the synthetic data, the methodology presented here was used to compute a harmonic mode representation of stress.  
In brief, the weights of the higher modes are determined through an optimization routine that
seeks to minimize violations of equilibrium.  
Both the grain boundary equilibrium violations and the intra-grain local equilibrium violations are built 
into the objective function.  
Their relative weighting will influence the optimized stress distributions. 
These weights are designated when assembling the complete objective function.  
Two observations are worth noting relative to the selection of ($w^b$,$w^v$):  (1) the limiting case of (0,1) does not provide any change from input data because a constant stress within grains satisfies equilibrium identically;  (2) the limiting case of (1,0) does induce changes (activate modes) but induces unreasonably large variations in the stress over the grains.  
Consequently,  pairs with non-zero values of ($w^b$,$w^v$) give more reasonable results, especially with a bias toward $ w^v > w^b$. 
In this example, ($w^b$,$w^v$) $=$ (0.03, 1.0).  
No attempt has been made to optimized this choice.

\begin{figure}[htbp]
	\centering
	\begin{subfigure}{.30\textwidth}
		\centering
		\includegraphics[width=1\linewidth]{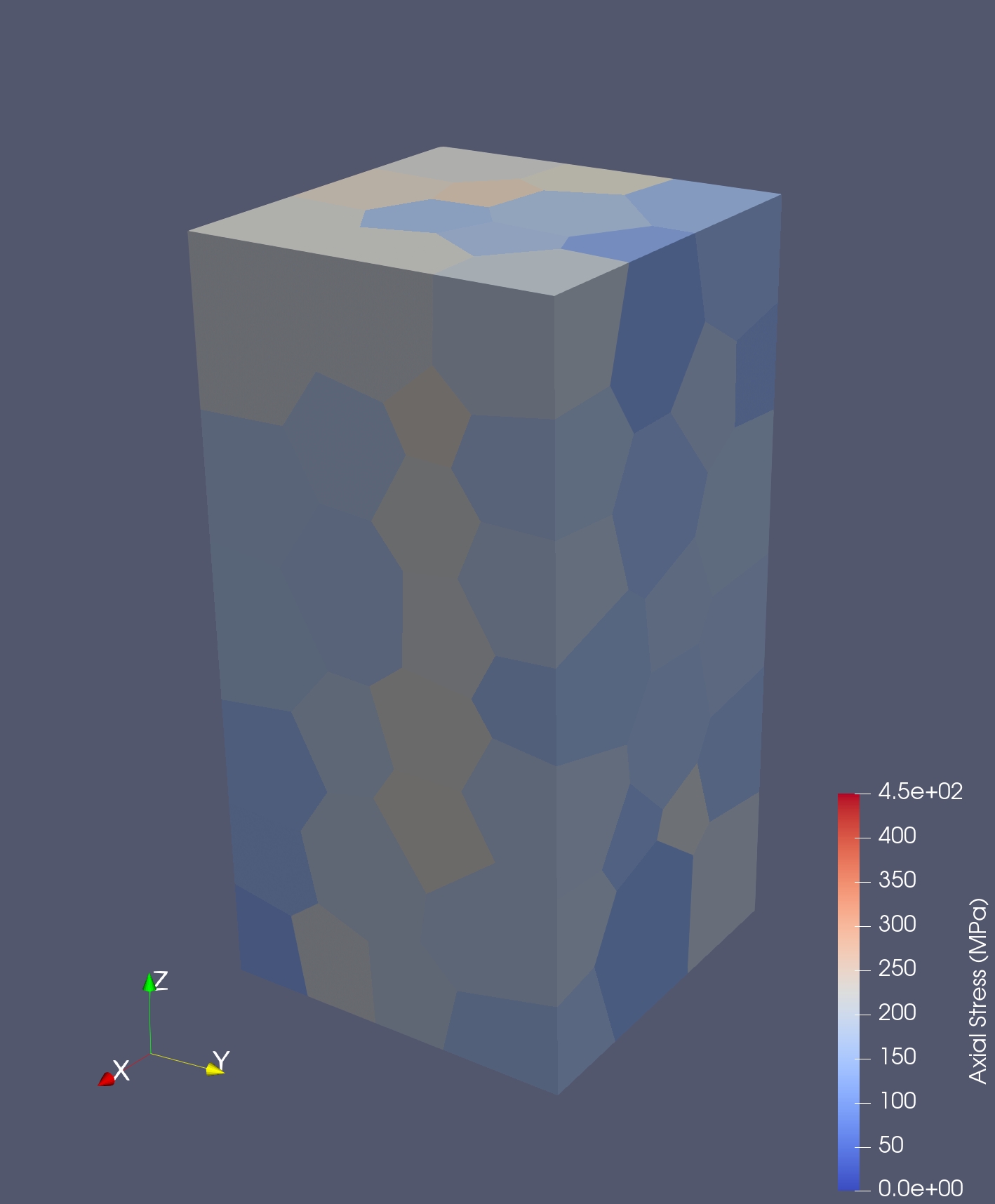}
		\caption{ }
		\label{fig:sr_1mode_sigzz}
	\end{subfigure}%
	\quad
	\begin{subfigure}{.30\textwidth}
		\centering
		\includegraphics[width=1\linewidth]{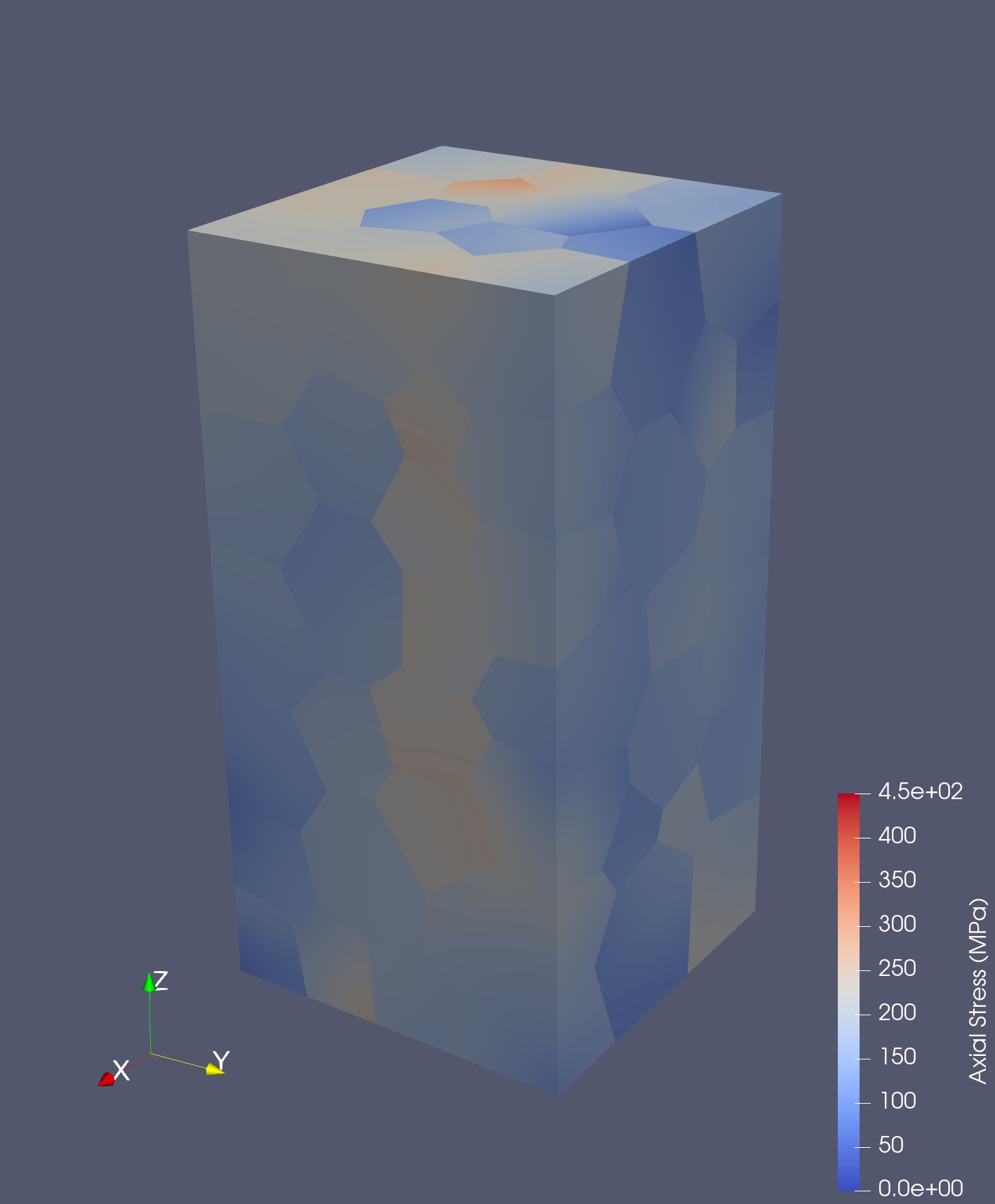}
		\caption{ }
		\label{fig:sr_4mode_sigzz}
	\end{subfigure}\\
	\begin{subfigure}{.30\textwidth}
		\centering
		\includegraphics[width=1\linewidth]{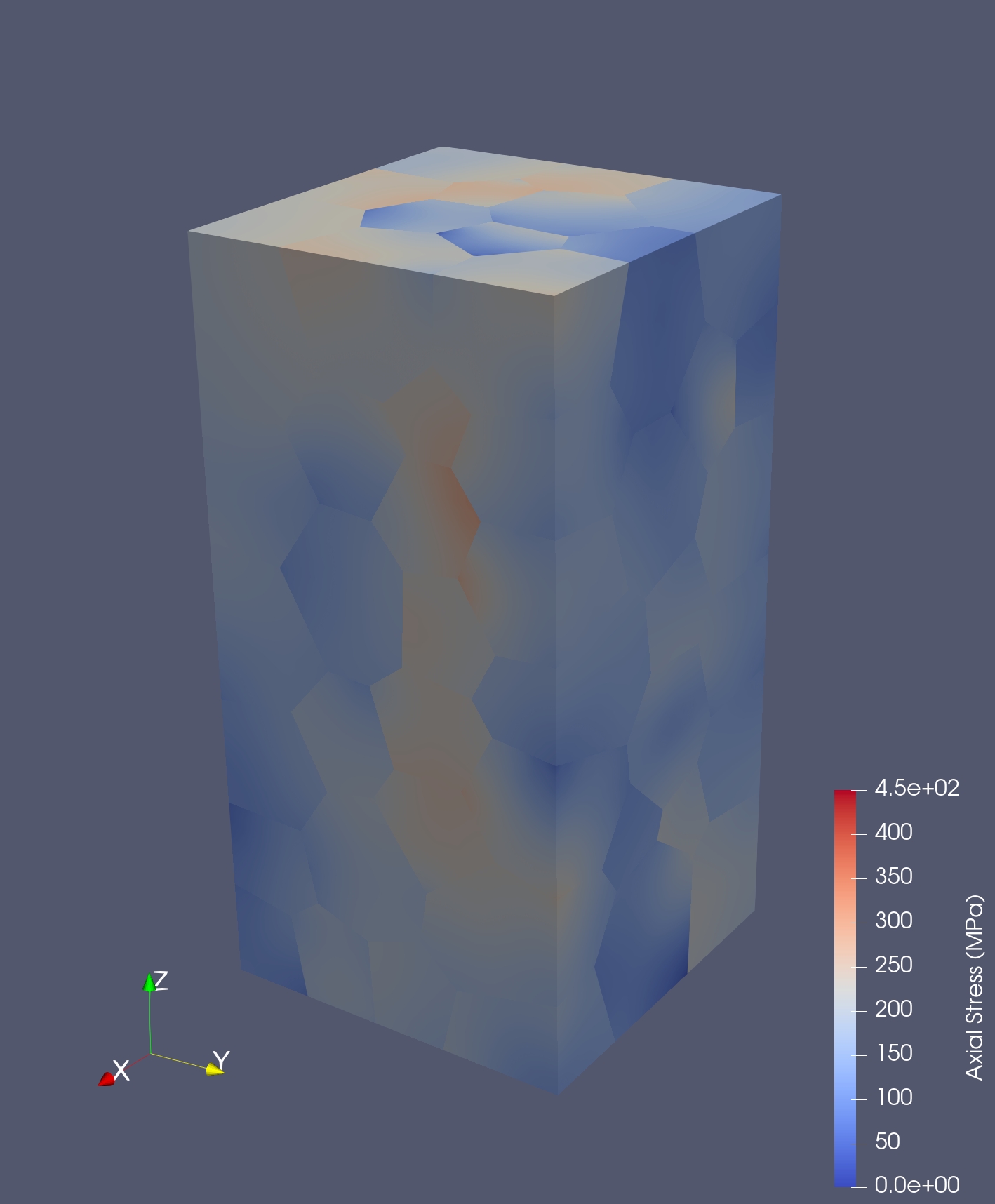}
		\caption{ }
		\label{fig:srt_10mode_sigzz}
	\end{subfigure}%
	\quad
	\begin{subfigure}{.30\textwidth}
		\centering
		\includegraphics[width=1\linewidth]{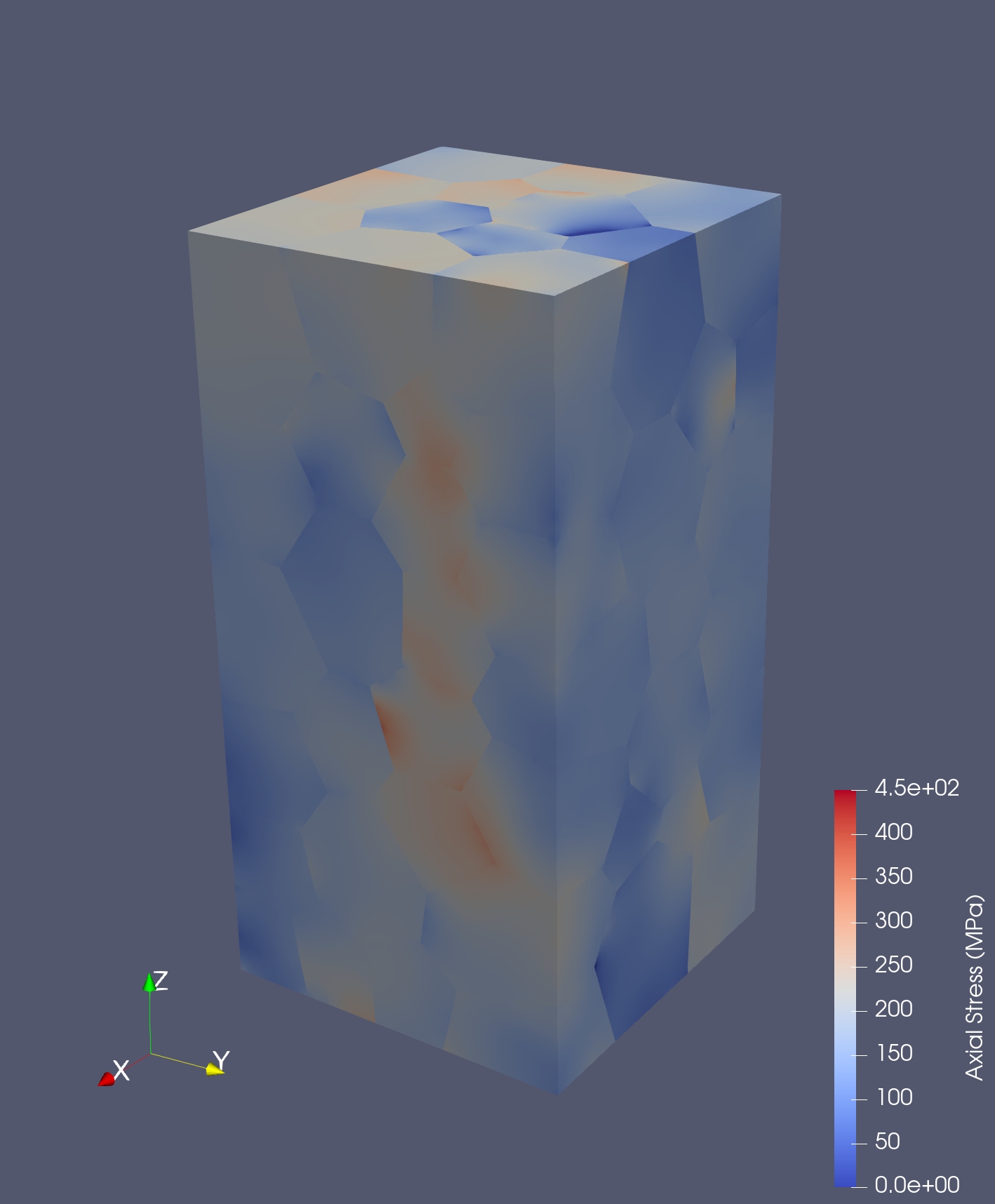}
		\caption{ }
		\label{fig:sr_28mode_sigzz}
	\end{subfigure}%
	\caption{Axial stress distributions from the reconstructed stress distributions for: (a) 1 mode; (b) 4 modes; (c) 10 modes; and (d) 28 modes. }
		\label{fig:sample_axial_stress_stress_recovery}
\end{figure}

With the values of $w^b$ and $w^v$ specified, the 
 reduction of the violations with increasing numbers of modes was examined
 to assess the sensitivity of the equilibrium violations to the intra-granular stress
 variations. 
To this end, analyses using expansions with 1, 4, 10 and 28 modes ($ 1, (3^1+1),( 3^2+1) {\, \rm and \,} (3^3+1)$)
were conducted and compared.
Figure~\ref{fig:sample_axial_stress_stress_recovery} shows the axial stress distributions corresponding
to expansions with differing numbers of modes.
The improvement evident from visual inspection of the four distributions is confirmed by the decreasing values of the objective function, $F$, given in Table~\ref{tab:sr_converg_stats}.  
Here, the values are normalized by the value of $F$ for 1 mode, equivalent to the input data. 
These diminish monotonically with increased number of modes.  
For 28 modes, the value of $F$ is less than 20\% of the value of $F$ for the grain-averaged stress (which is equivalent to the 1 mode expansion).  
Note that the discontinuities of axial stress at grain boundaries with normal vectors aligned with loading direction diminish with higher numbers of modes.
For these boundaries,  the axial stress dominates the traction vector.  
Thus, the discontinuities of axial stress essentially are the imbalances of tractions and are what is quantified in the objective function.  
\begin{table}[ht]
\small
\centering
\begin{tabular}{|c | c | c |}
\hline
    Modes $(n)$	& $F (\rm{MPa})^2$ & $F_n / F_1$ \\ 
\hline
1 & 24.40e6& -	\\
4& 15.10e6& 0.619	\\
10 & 9.47e6& 0.388	\\
28& 4.58e6& 0.188	\\
 \hline
\end{tabular}
\caption{Relative values of objective function for increasing numbers of expansion modes for ($w^b$,$w^v$) $=$ (0.03, 1.0).}
\label{tab:sr_converg_stats}
\end{table}

In Table~\ref{tab:expansion_weights_gr49_sigzz} the weights are tabulated for the axial stress component of Grain 49.  
Grain 49 is the right grain in Figure~\ref{fig:traction_constraint_2} and is centrally located in the sample..  
The  finite element stress field and the fields determined by fitting harmonic expansions to the finite element field are discussed in detail in \cite{Mechmonics_2021}.
Also provided in Table~\ref{tab:expansion_weights_gr49_sigzz} are the weights for the fitted distribution.
Several observations are possible for the reconstructed distribution weights: (1) the weight of Mode 1 is fixed at the input value as a consequence of the equality constraint; (2) as more modes are used, the weights of lower, repeated modes change -- sometimes increasing and other times decreasing; (3) higher modes have on average lower weights than lower modes. 
Comparing the fitted and reconstructed weights it is clear that the reconstructed distributions are not converging on the fitted distribution.  
\begin{table}[ht]
\small
\centering
\begin{tabular}{|c | c | c | c | c | c | }
\hline
    Mode Number & Fitted	& 1 Mode & 4 Modes & 10 Modes & 28 Modes\\ 
\hline
1& -1.0029 & -1.0029 &-1.0029 & -1.0029 & -1.0029	\\
2 & -0.0865 & - &-0.0271& -0.0377 & -0.0381	\\
3 & 0.0533& - &0.0089 & 0.0129 & 0.0025	\\
4 & 0.0751 & - &0.0039 & 0.0220 & 0.0081\\
5 & -0.0015 & - &- & 0.0048 & -0.0238	\\
6 & -0.0265 & - &- & -0.0059 & -0.0267\\
7 & -0.0321 & - &- & 0.0274 & -0.0109	\\
8 & -0.0090 & - &- & 0.0049& -0.0077	\\
9 & 0.0066 & - &- & -0.0138 & -0.0045	\\
10 & 0.0189 & - &- & 0.0090 & 0.0152\\
11& -0.0216 & - &- & - & -0.0050	\\
12& -0.0160 & - &- & - & -0.0088	\\
13& 0.0084 & - &- & - & -0.0010	\\
14& -0.0195 & - &- & - & 0.0082	\\
15& 0.0185 & - &- & - & 0.0082	\\
16& 0.0249 & - &- & - & -0.0028	\\
17& 0.0196 & - &- & - & 0.0098	\\
18& -0.0154 & - &- & - & 0.0021	\\
19& 0.0060 & - &- & - & -0.0114	\\
20& -0.0158 & - &- & - & -0.0090	\\
21& -0.0112 & - &- & - & -0.0025	\\
22& -0.0075 & - &- & - & -0.0005	\\
23& 0.0187 & - &- & - & 0.0033	\\
24& 0.0010 & - &- & - & 0.0009\\
25& 0.0020 & - &- & - & 0.0018	\\
26& -0.0166 & - &- & - & -0.0060	\\
27& 0.0071 & - &- & - & 0.0035	\\
28& -0.0041 & - &- & - & -0.0030	\\
 \hline
\end{tabular}
\caption{Expansion weights for fitted and recovered axial stress distributions in Grain 49. All weights have been divided by $10^4$.}
\label{tab:expansion_weights_gr49_sigzz}
\end{table}

\section{Discussion}
\label{sec:discussion}

\subsection{Comparison of original and reconstructed stress distributions } 
While there exist many similarities in the fitted and reconstructed stress distributions, there are differences
that persist with refinement.  
In general,  the stress distribution reconstructed from an averaged field by minimizing equilibrium violations
is not expected to recover the original field.
The reasons for this can be explained by considering the attributes of the two fields.
A finite element  solution with displacement (or velocity, for rate formulations) as the primary field variable has several defining attributes.  
First, the displacement field is determined from a set of possible fields that all satisfy compatibility.   
The  displacement fields are smooth and the strains are computed from the smooth displacement field. 
Second, the stresses are determined from the strains using Hooke's law (in the case of linear elastic behaviors).   Third, equilibrium is used to determine which displacement field from the set of possible ones via a residual over the entire volume together with surface tractions.   
Thus, all of the field equations enter the solution and, for elasticity, the finite element solution is known to converge on the exact solution with mesh refinement.

In contrast, the optimization procedure used in the stress reconstruction method quantifies equilibrium violations at grain boundaries and within grain interiors and seeks to minimize these among candidate stress distributions (ones possible with the harmonic expansion by varying the mode weights). 
There is no consideration of strain compatibility that is embedded in the finite element solution 
through the specification of a smooth displacement field.

Strain compatibility at the grain boundaries merits particular attention for polycrystals of interest here.  
The finite element stress distribution is determined from displacement fields that enforce straining in adjoining grains to be the same for strain components associated with in-plane deformation (grain boundaries stretch and shear identically in the shared interface).  
There is no special attention given to tractions across the grain boundaries as equilibrium is considered locally over the grain volumes.  
In contrast, the reconstructed stress methodology does penalize violations of traction equilibrium at grain boundaries. 
However, there is no penalty associated with violation of compatibility and therefore violations of compatibility will in all likelihood exist if strains were to be computed from the reconstructed stress fields. 
Because the reconstructed stress is not constrained to satisfy compatibility in the rigorous sense that the finite element solution is, there is no expectation that the reconstructed stress will converge to the finite element distribution.  
Rather, it will utilize the available degrees of freedom (weights of the expansion modes) to best match grain boundary tractions without regard to the consequences for compatibility.
 Another way of expressing this is that compatibilty restricts the admissible solutions available to the finite element method to a different set than does the reconstruction method that is based on equilibrium alone (whether done locally or both locally and at grain boundaries, as is done here).

\subsection{Alternative approaches}
Not enforcing compatibility in the reconstruction process leads to optimal (as defined by the equilibrium-based objective function) distributions that differ from the original. 
This motivates a question regarding whether it might improve the result by focusing on compatibility instead of or in addition to equilibrium for the objective function.
To this end, the objective function could be written to include strain compatibility at grain boundaries.  

One approach could be to completely overhaul the approach to a strain-based formulation:
(1) assume grain-averaged strain tensors are the data provided; 
(2) write the strain field in terms of harmonic expansion with unknown weights; 
(3) find the weights by minimizing an objective function that quantifies strain mismatch within grain boundaries and compatibility of strain over the grain volumes;  
and (4) enforce the grain-averaged data via an equality constraint. 
A second approach could be to modify the existing stress-based approach by adding grain boundary compatibility violations to the objective function. 
In this case the strains would be written in terms of the stresses using Hooke's law so that the structure of the matrix form of the objective function remains the same (still written in terms of the unknown weights on the expansion modes for the stress).
The former option most likely would be limited to elastic deformations because HEDM is only capable of quantifying the lattice (elastic) strain tensors. 


\section{Future work}
A number of possibilities exist for further developing and using the method.  
Two that have a high priority are:
\begin{itemize}
\item Investigate the possibility of extending the objective function and/or  constraints to penalize violations of compatibility (limited to elastic deformations).  
\item Investigate the application of the method to the evaluation of residual stresses in engineering components (such as additive manufacturing) for initialization of stress in modeling.
\end{itemize}

\section{Summary}

The methodolgy provides a means to augment grain-averaged stress distributions with intra-grain variations.
The method formulates an equilibrium-based objective function in terms of expansion functions and 
evaluates the expansion weights by minimizing equilibrium violations  at grain boundaries and within grain interiors.
The method is demonstrated for a case with ``synthetic'' x-ray data generated by a \mechmet\, simulation.
The synthetic data define a grain-averaged stress field in a virtual polycrystal that was
subjected to axial extension.  
Using the method, intra-grain variations were added to the distribution by means of
weighted harmonic modes.  
With increasing numbers of modes, there is a monotonic decrease in the equilibrium violation
as quantified by the values of the objective function.  
An important observation made from comparing the reconstructed and original distributions was that the
reconstructed stress distribution did not converge on the stress distribution obtained by evaluating the harmonic weights directly from the original stress distribution.
This point is discussed in terms of  attributes of the original and reconstructed distributions
imposed by the differences in which equilibrium and compatibility are imposed.

\section*{Acknowledgements}
The research reported here was supported by the ONR under grant \# N00014-16-1-3126, Dr. William Mullins Program Manager.

%
 \section*{Conflict of interest}
 The authors declare that they have no conflict of interest.

\bibliographystyle{spmpsci}      
\bibliography{Stress_Recovery.bib}  

\end{document}

%% file: mymathmacros.tex
\DeclareMathAlphabet{\mathsfsl}{OT1}{cmss}{m}{sl}

\newcommand{\fepx}{{\bfseries{\slshape{FEpX}}}}
\newcommand{\mechmet}{{\bfseries{\slshape{MechMet}}}}
\newcommand{\mechmonics}{{\bfseries{\slshape{MechMonics}}}}

\newcommand{\neper}{{\bfseries{\slshape{Neper}}}}

\newcommand{\matlab}{{\bfseries{\slshape{MATLAB}}}}

\newcommand{\vctr}[1]{\boldsymbol{#1}}

\newcommand{\tnsr}[1]{\mathsfsl{#1 }}

\newcommand{\cauchy}{{\boldsymbol{\sigma}}}

\newcommand{\dee}{{\mathrm{d}}}